\documentclass[12pt,a4paper]{article}
\pdfoutput=1
\usepackage{cite,color,graphics,amsmath,epsfig,rotating}
\usepackage{latexsym}
\usepackage{amssymb}

\usepackage{graphicx}
\usepackage{slashed}
\usepackage{subfig}
\usepackage{psfrag}
\usepackage{mathrsfs}
\usepackage{url}
\ifx\pdfoutput\undefined
\usepackage[bookmarks]{hyperref}	
\usepackage{enumitem}
\else
\usepackage{hyperref}	
\fi
\hypersetup{colorlinks=true,linktocpage,bookmarksopen,bookmarksnumbered,citecolor=azur,
linkcolor=green1,pdfstartview=FitH,urlcolor=darkred,breaklinks=true,
pdftitle={micrOMEGAs_5.0 : freeze-in},
pdfauthor={G. B\'elanger, F. Boudjema, A. Goudelis, A. Pukhov, B. Zaldivar},
pdfsubject={paper},
pdfkeywords ={beyond the standard model, new physics, dark matter, Higgs, LHC}}

\usepackage{multicol}
\usepackage{color}
\definecolor{azur}{rgb}{0.118,0.498,0.796}
\definecolor{darkred}{cmyk}{0,1,1,0.4}
\definecolor{green1}{rgb}{0.21,0.6,0.32}
\def\mhref#1{\href{mailto:#1}{#1}}		

\usepackage{marginnote}

\newcommand{\beq}{\begin{equation}}
\newcommand{\eeq}{\end{equation}}
\newcommand{\bea}{\begin{eqnarray}}
\newcommand{\eea}{\end{eqnarray}}

\usepackage{eurosym}
\begin{document}

\setlength{\unitlength}{1mm}
\renewcommand{\arraystretch}{1.4}
\newcommand{\comment}[1]{}
\newcommand{\gb}{\textcolor{blue}}
\newcommand{\ag}{\textcolor{magenta}}


\def\micro{{\tt micrOMEGAs}}
\def\microvn{{\tt micrOMEGAs\,4.3}}
\def\wimpsim{{\tt WimpSim}}
\def\pppc{{\tt }PPPC4DM$\nu$}
\def\pppcold{{\tt }DM$\nu$}
\def\chep{{\tt CalcHEP}}
\def\lhep{{\tt LanHEP}}
\def\darksusy{{\tt DarkSUSY}}
\def\smodels{{\tt SModelS}}
\def\superiso{{\tt SuperIsoRelic}}
\def\maddm{{\tt MadDM}}
\def\madanalysis{{\tt MadAnalysis\,5}}
\def\checkmate{{\tt CheckMate}}
\def\nmssmtools{{\tt NMSSMTools}}
\def\fastlim{{\tt Fastlim}}
\def\xqcat{{\tt XQCAT}}
\def\lilith{{\tt Lilith}}
\def\HB{{\tt HiggsBounds}}
\def\HS{{\tt HiggsSignals}}
\def\suspect{{\tt SuSpect}}
\def\neuto{\tilde\chi^0_1}
\def\neuti{\tilde\chi^0_i}
\def\neutt{\tilde\chi^0_2}
\def\neuth{\tilde\chi^0_3}

\def\ie{{\it i.e.}}
\def\eg{{\it e.g.}}

\def\br{{\rm BR}}

\def\ra{\rightarrow}
\def\Omg{\Omega h^2}
\def\sip{\sigma^{SI}_{\chi p}}
\newcommand{\scs}{\scriptscriptstyle}
\def\simleq{\stackrel{<}{\scs \sim}}
\def\simgeq{\stackrel{>}{\scs \sim}}
\newcommand{\com}{\textcolor{green}}
\newcommand{\combis}{\textcolor{magenta}}

\newcommand{\ablabels}[3]{
  \begin{picture}(100,0)\setlength{\unitlength}{1mm}
    \put(#1,#3){\bf (a)}
    \put(#2,#3){\bf (b)}
  \end{picture}\\[-8mm]
}

\begin{titlepage}
\begin{center}

\vspace*{1.6cm}
{\Large\bf micrOMEGAs5.0 : freeze-in} 

\vspace*{1cm}\renewcommand{\thefootnote}{\fnsymbol{footnote}}

{\large 
G.~B\'elanger$^{1}$\footnote[2]{Email: \mhref{belanger@lapth.cnrs.fr}},
F.~Boudjema$^{1}$\footnote[3]{Email: \mhref{boudjema@lapth.cnrs.fr}},
A.~Goudelis$^{2}$\footnote[4]{Email: \mhref{andreas.goudelis@lpthe.jussieu.fr}}, 
A.~Pukhov$^{3}$\footnote[5]{Email: \mhref{pukhov@lapth.cnrs.fr}},
B.~Zaldivar$^{1}$\footnote[8]{Email: \mhref{zaldivar@lapth.cnrs.fr}} 

\renewcommand{\thefootnote}{\arabic{footnote}}

\vspace*{1cm} 
{\normalsize \it 
$^1\,$ Univ. Grenoble Alpes, USMB, CNRS, \href{http://lapth.cnrs.fr}{LAPTh}, F-74940 Annecy, France\\[2mm]
$^2\,$ Sorbonne Universit\'e, CNRS, Laboratoire de Physique Th\'eorique et Hautes \'Energies, \href{http://lpthe.jussieu.fr/}{LPTHE}, F-75252 Paris, France\\[2mm]
$^3\,$\href{http://theory.sinp.msu.ru}{Skobeltsyn Institute of Nuclear Physics}, Moscow State University,\\ Moscow 119992, Russia\\[2mm]
}}

\vspace{1cm}

\begin{abstract}
We present a major upgrade of the micrOMEGAs dark matter code to compute the abundance of feebly interacting dark matter candidates through the freeze-in mechanism in generic extensions of the Standard Model of particle physics. We develop the necessary formalism in order to solve the freeze-in Boltzmann equations while making as few simplifying assumptions as possible concerning the phase-space distributions of the particles involved in the dark matter production process. We further show that this formalism allows us to treat different freeze-in scenarios and discuss the way it is implemented in the code. We find that, depending on the New Physics scenario under consideration, the effect of a proper treatment of statistics on the predicted dark matter abundance can range from a few percent up to a factor of two, or more. We moreover illustrate the underlying physics, as well as the various novel functionalities of micrOMEGAs, by presenting several example results obtained for different dark matter models.
\end{abstract}

\end{center}
\end{titlepage}

\tableofcontents

\section{Introduction}

In the last decades, dark matter (DM) studies have mainly focused on frozen-out weakly interacting massive particles (WIMPs). This tendency has, in part, been motivated by theoretical arguments: on one hand, thermal freeze-out is a phenomenon that is known to occur for several other particle species in the early Universe. On the other hand WIMPs, \textit{i.e.} particles with masses and interaction strengths lying within a few orders of magnitude around the electroweak scale, are ubiquitous in numerous well-motivated New Physics scenarios and in particular in models trying to address the hierarchy problem. The fact that WIMP freeze-out naturally leads to the dark matter abundance inferred in particular from the Cosmic Microwave Background \cite{Hinshaw:2012aka,Ade:2015xua} has, thus, corroborated them as plausible dark matter candidates. The interest in WIMPs is, nonetheless, not only due to theoretical prejudice. WIMP freeze-out is typically driven by non-negligible interactions between the visible and the ``dark'' sector. The same interactions could also mediate DM production at high-energy colliders, scattering of DM off ordinary matter and, eventually, DM annihilation in the galaxy and beyond. Furthermore, in numerous WIMP constructions DM is only the lightest stable state of an entire family of particles that can interact with the standard model (SM) with comparable strengths. These features imply that WIMP scenarios can give rise to observable signals. The WIMP freeze-out picture, hence, motivated an impressive experimental search programme worldwide in order to constrain and, eventually, unravel such non-gravitational properties of dark matter, for reviews see \eg~\cite{Bertone:2004pz,Bertone:2010zza}. 

However, the stringent bounds that have been placed on WIMP dark matter candidates from direct detection~\cite{Aprile:2017iyp,Tan:2016zwf,Akerib:2016vxi,Amole:2017dex}, indirect detection~\cite{Ahnen:2016qkx,Abdallah:2016ygi,Giesen:2015ufa} and collider searches~\cite{Aad:2015baa,Khachatryan:2016nvf} have revived the interest in candidates with much different interaction strengths. In particular, a dark matter particle that is feebly coupled with the standard model, that is typically with a coupling strength ${\cal O} (10^{-10}-10^{-12})$, easily escapes the bounds from standard WIMP searches, yet can reproduce the precisely measured value for the DM density~\cite{Kim:2017mtc,Kaneta:2016wvf,Heikinheimo:2016yds,Ayazi:2015jij,Molinaro:2014lfa,Blennow:2013jba,Dev:2013yza,Chu:2013jja,Klasen:2013ypa,Mambrini:2013iaa,Yaguna:2011qn,Chu:2011be,Shakya:2015xnx,Merle:2015oja,Nurmi:2015ema,Pandey:2017quk,Arcadi:2016dbl,Benakli:2017whb}. This can be achieved through the so-called ``freeze-in'' mechanism, where the feebly interacting massive particle (FIMP) cannot reach thermal equilibrium in the early universe and is produced either from the scattering or from the decay of particles in the thermal bath~\cite{McDonald:2001vt,Hall:2009bx}. Contrary to the case of freeze-out, where the predicted DM relic density is inversely proportional to its thermally averaged self-annihilation cross-section, in freeze-in the two quantities scale proportionally. In many models the dominant contribution occurs at a temperature of the order of the DM mass or the mediator mass, therefore there is little dependence on the temperature at which DM production first starts as long as the latter is much higher than the other scales in the model. However, in some cases the production mechanism can be sensitive to high temperatures, \textit{e.g.} in scenarios involving non-renormalisable operators~\cite{Mambrini:2013iaa,Elahi:2014fsa,McDonald:2015ljz,Mambrini:2015vna} or even in renormalisable models in which the DM production cross section is constant at high energies~\cite{vll}.

The computation of the most precisely measured DM observable, its density in the Universe, is well-established for WIMPs.
It can be challenging in models with a large number of new particles, \textit{e.g.} in supersymmetry. For this reason a number of numerical codes have been developed including \micro~\cite{Barducci:2016pcb,Belanger:2006is,Belanger:2001fz}, \darksusy~\cite{Gondolo:2004sc}, \superiso~\cite{Arbey:2011zz} and {\tt MadDM}~\cite{Backovic:2013dpa}. The corresponding Boltzmann equations for FIMPs are also known and have been solved in several models, for a review see also~\cite{Bernal:2017kxu}. However, to this date there exists no public computational tool to perform this task.
This is the gap we intend to fill here. 

Although, as will become apparent in the following (\textit{cf} also \cite{Hall:2009bx,Bernal:2017kxu}), the general thermodynamics governing DM genesis is shared amongst different freeze-in scenarios, in practice it is useful to distinguish between several possibilities depending on the nature, the mass and the interaction strength of the degrees of freedom involved in the mechanism.
First, some states may only couple to pairs of DM particles whereas others may share some quantum numbers with dark matter. 
Following the classification we have introduced in previous versions of \micro~we will implicitly assume that all particles are 
either ``even'' or ``odd'' with respect to some discrete symmetry. Odd particles, including the DM relic, form the Dark Sector (DS). Even particles, including the SM degrees of freedom, with direct couplings to both SM particles and DS pairs will be referred to as ``mediators''\footnote{Note however that in order to comply with commonly used terminology, and unless this might lead to confusion, we will sometimes collectively refer to both even and odd particles other than DM itself as ``mediators''.}. Secondly, depending on the strength of their interactions, mediators may or may not be in thermal equilibrium with the SM. The same statement applies to the DS particles.
This leads us to further divide particles into two more classes: class ${\cal F}$ contains feebly interacting particles and class ${\cal B}$ (the bath) contains all particles that are in thermal equilibrium with the SM. Then, in a freeze-in scenario, the DM particle is the lightest ``odd'' state that belongs to ${\cal F}$. One of the simplest cases is the one in which the (even) mediator is in thermal equilibrium with the SM and can decay, with a small branching ratio, into a pair of  DM particles. When the mediator is too light to decay into DM, the main relevant process is instead the $2\to 2$ annihilation of SM particles into DM. This process also dominates when the mediator is more weakly coupled to the SM than to DM. Another well-studied case is the one in which the DM is feebly coupled to some other odd  particle, the next-to-lightest dark sector particle (NLDSP), with the latter being a WIMP (in ${\cal B}$), for example  the gravitino, sneutrino  or axino in supersymmetric models with a neutralino next-no-lightest supersymmetric particle~\cite{Hall:2009bx}. In this case the DM abundance is simply related to the NLDSP one, which can be calculated according to standard thermal freeze-out. Other cases include the potential freeze-in of the NLDSP~\cite{Merle:2013wta} or the mediator themselves, a scenario which necessitates the simultaneous resolution of two coupled Boltzmann equations.

In this paper, we present a major extension of the \micro~dark matter code through which the user can, henceforth, consistently compute the freeze-in abundance of FIMP dark matter candidates. Starting from a very small initial number density, FIMPs are created during the thermal evolution of the universe from bath particle interactions - either through scattering processes or from the decay of a mediator or dark sector particle which may or may not be in thermal equilibrium with the SM. Within assumptions and limitations that will be discussed in detail in the following sections, the code works equally well for simple as well as more complicated New Physics scenarios, potentially containing multiple mediators and/or dark sector particles, as long as the DM is a FIMP. It can also compute the freeze-in abundance of a FIMP that subsequently decays into WIMP DM. 

Besides providing a numerical code that can work with many extensions of the Standard Model, a novel aspect of the DM abundance calculations presented here is that we include a proper treatment of the phase-space distribution of bath particles rather than just assume a Maxwell-Boltzmann distribution as is usually done in the literature. Indeed, since DM production can start at high temperatures, the Fermi-Dirac or Bose-Einstein nature of the corresponding distributions has to be taken into account. We find that the inclusion of these effects can lead to a factor two variation in the predicted value of the relic density, especially when it comes to annihilation processes initiated by bosons. We exemplify the results obtained through \micro~by computing the FIMP freeze-in abundance within a few simple models: a model with a $Z'$ mediator and a Dirac fermion DM, the minimal singlet scalar DM model and a model containing an additional dark sector particle and which has interesting phenomenological implications for long-lived particle searches at the LHC and beyond.
\\
\\
This new version of \micro~includes new functionalities for the computation of the relic density of DM through freeze-in while keeping all the previous features for  DM freeze-out, specifically \\ 
\begin{itemize}
\item{} New functions to compute the relic density of FIMPs. The user must explicitly specify which particles are to be considered as FIMPs. The reheating temperature, that is the temperature at which dark matter formation starts, must also be  specified by the user. The code currently supports freeze-in computations for up to two dark matter candidates.
\item{} A proper treatment of the phase-space distribution functions for  bosons and fermions  when computing the dark matter number density.
\item{} The possiblity to specify each process to be included in the computation of freeze-in cross sections in order to examine  the contribution of each  channel to the DM production. 
\item{} The possibility to compute the relic density of the lightest dark sector particle (LDSP) or the NLDSP via freeze-out.
\item{} Three new sample models in which DM production proceeds through freeze-in that can be used as examples for the incorporation of other New Physics scenarios. 
\end{itemize}
Note also that within a model, the user can call routines to compute the relic density via freeze-in or routines that rely on the freeze-out picture. Unless some particles are explicitly declared as being feebly interacting, all freeze-out routines will work exactly like in previous releases of \micro. Specific signatures of FIMPs, through direct detection~\cite{An:2014twa}, indirect detection~\cite{Essig:2013goa,Heikinheimo:2018duk} and collider searches~\cite{Arcadi:2014tsa,Aaboud:2016uth,Khachatryan:2016sfv,Ghosh:2017vhe} will not be discussed here and are a topic of future work.

The paper is organised as follows. 
In Section~\ref{sec:relic} we present the equations governing the computation of the relic density of FIMPs and the way they are solved within \micro. 
In Section~\ref{sec:micro} we describe the \micro' functions relevant for freeze-in and their utilisation.
In Section~\ref{sec:models} we study a few sample models which are provided with the code as examples, and present some numerical results. 
Installation instructions as well as a sample output are given in Section~\ref{sec:install}.
Finally, section~\ref{sec:conclusion} contains our conclusions. 
Some more technical aspects concerning the treatment of $t$-channel propagators are presented in the Appendix.

\section{Relic density of FIMPs} 
\label{sec:relic}

When the coupling of DM to the SM sector is very weak, the two never reach thermal equilibrium in the  early Universe. Assuming that the initial number density of DM is very small it can, then, be produced from the decay or annihilation of other particles, typically until the number density of the latter becomes small due to Boltzmann suppression\footnote{This is, of course, a schematic picture. As will become apparent in the following, the DM production process can be more complicated, involving intermediate, potentially long-lived particles. Note also that contrary to the usual freeze-out picture, a larger FIMP coupling to the visible sector implies a larger DM density.}. Moreover, given the smallness of the DM couplings and number density, its annihilations into bath particles can be ignored. The DM abundance then freezes-in to a constant value which can meet the one inferred by CMB measurements.

\subsection{Notations and useful formulas}

Before presenting the equations governing freeze-in production of dark matter and our approach towards their resolution, let us begin by swiftly introducing some notations and remind the reader some useful formulas. 

Assume a particle species $i$ in the early Universe, obeying a phase space distribution function $f_i$. The number and energy density of the $i$ particles is given by
\begin{equation}
\label{eq:numberenergydensitydef}
n_i = \frac{g_i}{(2\pi)^3} \int f_i(\vec{p}) d^3\vec{p} 
\qquad {\rm and} \qquad 
\rho_i = \frac{g_i}{(2\pi^3)} \int E_i(\vec{p}) f_i(\vec{p}) d^3\vec{p}
\end{equation}
respectively, where $g_i$ is the number of internal degrees of freedom of the species, $\vec{p}$ is the $3-$momentum and $E_i$ is the energy. Particles in kinetic equilibrium with a thermal bath follow the usual Fermi-Dirac/Bose-Einstein distributions which, in the absence of any anisotropies, take the usual form
\begin{equation}
\label{eq:FD-BE}
f_i = \frac{1}{[\exp(E_i-\mu_i)/T\pm 1]} =  \frac{\left| \eta_i \right|}{e^{\frac{E_i}{T}} -\eta_i} 
\end{equation} 
where $\mu_i$ is the chemical potential. In the second step we have introduced the useful parameter $\eta_i \equiv \eta_s e^{\mu_i/T}$, where $\eta_s= 1$ for bosons, $\eta_s= -1$ for fermions and $\eta \approx 0$ for FIMPs.
\\
\\
For $T \gtrsim 100$ eV, \textit{i.e.} in the radiation dominated Universe, the energy and entropy density can be written as a function of the corresponding effective number of degrees of freedom $g_{\rm eff}$ and $h_{\rm eff}$ as
\begin{align}
\label{geff}
      \rho(T) & = \frac{\pi^2}{30} T^4  g_{\rm eff}(T) \\
\label{heff}
     s(T) & = \frac{2\pi^2}{45} T^3 h_{\rm eff}(T)~.
\end{align}
The Hubble expansion rate is defined as
\begin{equation}
\label{Hubble}
  H(T)=\frac{1}{M_{Pl}} \sqrt{\frac{8\pi}{3}\rho(T)}
\end{equation}
where $M_{Pl}=1.22\cdot10^{19}$ GeV is the Planck mass. For a rough estimate  $H(T) \approx 10^{-18} (T/{\rm GeV})^2$ GeV. Finally, entropy conservation (which amounts to $ds/dt=-3Hs$) allows us to relate time and temperature through 
\begin{equation}
\label{eq:timetemperaturerelation}
dt = -\frac{dT}{\overline{H}(T) T}
\end{equation}
with the function $\overline{H}(T)$ given by 
\begin{equation}
\label{eq:Hprimedef}
\overline{H}(T) = \dfrac{H(T)}{1+\frac{1}{3}\frac{d \ln \left( h_{\rm eff}(T) \right)}{d \ln T}}.
\end{equation}
 
\subsection{General freeze-in Boltzmann equation}\label{sec:generalFI}

The time/temperature evolution of the number density of a particle species $\chi$ in the early Universe can be described by a Boltzmann equation which, for a Friedmann-Lema\^itre-Robertson-Walker Universe, reads in full generality
\begin{equation}
\label{eq:GeneralBoltzmann}
\dot{n}_\chi + 3Hn_\chi = \sum_{A,B} ( \xi_{B} - \xi_{A}) {\cal N} (A \rightarrow B)
\end{equation}
where $A$ and $B$ denote generic initial and final states containing $\xi_{A,B}$ particles of type $\chi$ respectively and ${\cal N} (A \rightarrow B)$ is the integrated collision term corresponding to the reaction $A \rightarrow B$, \textit{i.e.} the number of $A\rightarrow B$ reactions taking place in the thermal bath per unit space-time volume.

The integrated collision term for a process describing a set of particles $A$ going to a set of particles $B$ can, in turn, be written as
\begin{align}
{\cal N}(A\to B) & =  \int \prod_{i\in A} \left(\dfrac{d^3p_i}{(2\pi)^32E_i} f_i\right) 
\prod_{j\in B} \left(\dfrac{d^3p_j}{(2\pi)^32E_j} (1\mp f_j)\right)\nonumber\\
&\times (2\pi)^4\delta^4(\sum_{i\in A}P_i - \sum_{j\in B}P_j)C_{A}|{\cal M}|^2
\label{coll1}
\end{align}
where $C_{A}$ is a combinatorial factor which, focusing on $2 \rightarrow 2$ reactions, equals 1/2 in case of identical incoming particles and 1 otherwise, $P_i$ are the 4-momenta of the particles involved in the reaction, $f_i$ are their distribution functions and $|{\cal M}|^2$ is the squared transition matrix element summed over initial and final polarisations. 
\\
\\
In passing, let us point out that for final state particles, we can also rewrite
\begin{equation}
\prod\limits_{j \in out} (1\mp f_j) = \frac{e^{E_+/T}}{\prod\limits_{j \in out} [e^{E_j/T} - \eta_j]}  
\end{equation}
where $E_+=\sum_j E_j$. Then, because of energy conservation, Eq.~\eqref{coll1} is symmetric with respect to the interchange $A\leftrightarrow B$ except for a global factor $e^{-\sum_j \mu_j/T}$, therefore, 
\begin{equation}
{\cal N}(B\to A) = \dfrac{\prod_{j \in B}| \eta_j|}{\prod_{i \in A} |\eta_i|} \dfrac{C_{B}}{C_{A}}{\cal N}(A\to B)~.
\label{collsym}
\end{equation}
These relations will be useful in the following.
\\
\\
The basic premises of the freeze-in scenario are that dark matter particles, hereafter denoted by $\chi$, are characterised by very small (``feeble'') couplings with the visible sector and by a negligible initial abundance, $f_\chi\ll 1$. 
These two assumptions allow us to set $\xi_A = 0$ in Eq.~\eqref{eq:GeneralBoltzmann}, \textit{i.e.} only consider DM production processes\footnote{In reality, the negligible initial abundance requirement is not necessary, as long as DM annihilation processes can be ignored: in this case, a non-zero initial abundance is simply an additive contribution to the freeze-in one. We will, nonetheless, make this assumption throughout our analysis.}.
For general values of the DM coupling strength to other particles in the spectrum, this assumption might or might not be valid. Assuming that DM is exactly stable, it constitutes a good approximation as long as
\begin{equation}
H(T) \gg n \langle \sigma v \rangle ~,
\end{equation}
where $\langle \sigma v \rangle$ is the thermally averaged DM annihilation cross-section times velocity. In \micro~we do \textit{not} check whether this condition holds or not.

Using the time/temperature relation \eqref{eq:timetemperaturerelation}, the final dark matter yield (\textit{i.e.} its comoving number density) $Y_\chi \equiv n_\chi/s$ can be obtained after integrating the collision term from the temperature at which DM production starts, which we will hereafter refer to as the \textit{reheating temperature} $T_R$, to the present temperature $T_0$ 
\beq
Y_\chi^0= \int\limits_{T_0}^{T_R} \dfrac{dT}{T\overline{H}(T)s(T)} \left( {\cal N}(bath \to \chi X)  +  2  {\cal N}(bath \to \chi \chi') \right)~,
\label{yieldDM}
\eeq
where $X$ stands for any bath particle and $\chi'$ for any dark sector FIMP, which we assume to (eventually) decay into a DM particle along with a visible sector one. We stress  that the abundance will depend on the value of the reheating temperature when this temperature is of the same order as the mediator or DM mass, or when the collision term is dominated by high temperatures. Note that in a given model there can also be channels that lead to the production of 3 or 4 DM particles, for example via the production of a pair of mediators that decay into DM. Such possibilities are not taken into account in the present version of \micro. The dark matter relic density is obtained using Eq.~\eqref{yieldDM}
\beq
\Omega h^2= \frac{m_\chi Y_\chi^0 s_0 h^2}{\rho_c}
\eeq
where the entropy density today is   $s_0= 2.8912 \times 10^{9}$ m$^{-3}$, $m_\chi$ is the dark matter mass and the critical density is $\rho_c= 10.537 \, h^2$ GeV m$^{-3}$. It is  related to the Hubble constant today, $H_0= h \,100$ km s$^{-1}$ Mpc$^{-1}$ with  $h=0.678(9)$.
\\
\\
As already alluded to in the introduction, we will distinguish amongst three situations which can  appear in  freeze-in scenarios:
\begin{enumerate}
\item{} dark matter production through $1\to 2$ decays of a mediator (an even particle decaying into a DM pair) or dark sector particle (an odd particle decaying into a DM and an even particle) in thermal equilibrium with the thermal bath,
\item{} dark matter production through $1\to 2$ decays of particles that are not in thermal equilibrium with the thermal bath (taking special care of the possibility of late decays) and
\item{} dark matter production through $2\to 2$ annihilations of a pair of bath particles. Here the final state can consist of either two DM particles or a DM and a bath particle.
\end{enumerate}
In everything that follows we always assume that the lightest particle of the odd sector is feebly interacting (FIMP).


\subsection{$1\to 2$ ($2\to 1$) processes}\label{sec:onetotwo}
Consider the case in which one or two DM particles are produced from the decay of a heavy particle $Y$, through a process of the type $Y \rightarrow a, b$, corresponding to scenarios 1 and 2 in the listing above. Following the terminology presented in the previous paragraph, $Y$ can be a SM particle, a new particle acting as a mediator that decays into two odd FIMPs ($a=\chi$, $b=\chi'$) or a dark sector particle that decays into one DM and one bath particle ($a=\chi, b={\rm bath}$). The formalism that we present here covers both possibilities. The integrated collision term \eqref{coll1} in this case obtains the form
\begin{align}
{\cal N}(Y\to a,b) & =  \int \dfrac{d^3p_Y}{(2\pi)^3 2E_Y} 
\dfrac{d^3p_a}{(2\pi)^32E_a} \dfrac{d^3p_b}{(2\pi)^32E_b} f_Y (1\mp f_a)(1\mp f_b) \nonumber\\
&\times (2\pi)^4\delta^4(P_Y - P_a - P_b)|{\cal M}|^2 \ .
\label{eq:coll12dec}
\end{align}
\subsubsection{Mathematical detour}
Before proceeding with the evaluation of the collision term, Eq.~\eqref{eq:coll12dec}, let us briefly digress in order to introduce a function that we will use heavily throughout the following. First, note that if we replace the distribution function $f_Y(p_Y)$ by $(2\pi)^3 \delta^3(\vec{p} - \vec{p}_Y)/g_Y$,which corresponds to one particle per unit volume, we can promptly perform the $d^3p_Y$ integration to obtain the quantity\footnote{We can exchange the role of $\vec{p}$ and $\vec{p}_Y$ within the integral in order to keep an explicit $Y$ index, which we find makes notations clearer.}
\begin{align}
G_{Y\to a,b}(p_Y)  & =  \frac{1}{2 E_Y} \int
\dfrac{d^3p_a}{(2\pi)^32E_a} \dfrac{d^3p_b}{(2\pi)^32E_b} (1\mp f_a)(1\mp f_b) \nonumber\\
&\times (2\pi)^4\delta^4(P_Y - P_a - P_b) \overline{|{\cal M}|}^2 \ .
\label{eq:YwidthInMedium}
\end{align}
In the limit $f_{a,b} \ll 1$, this expression corresponds to the usual (in-vacuum) decay rate $G_{Y\to a,b}$ of $Y$ into $a$ and $b$ in a reference frame in which $Y$ has an energy (3-momentum) $E_Y$ ($p_Y$) with respect to its rest frame (YRF). Then, including the factors from statistical quantum mechanics $(1\mp f_a)(1\mp f_b)$ in $G_{Y\to a,b}$, Eq.~\eqref{eq:YwidthInMedium} can be interpreted as the decay rate of $Y$ into $a$ and $b$ in the presence of the medium created by particles of this type. 

In this integral, the distribution functions $f_{a,b}$ are known in the comoving frame (CF). However, we know that the phase-space element is Lorentz-invariant. Then, we can write the latter in terms of YRF quantities while keeping the distribution functions in the CF, but rewriting all the relevant kinematic quantities in terms of their YRF counterparts, namely
\begin{align}
E_{a/b} \left(c_\theta\right) = E_{a/b}^{\rm YRF} \frac{E_Y}{m_Y} \pm \frac{|p_Y|}{m_Y} p_{a/b}^{\rm YRF} c_{\theta}
\end{align}
where $E_{a/b},E_Y$ and $p_Y$ are given in the CF while $c_{\theta}$ is the cosine of the angle between the momentum of particle $a$ and the Lorentz boost from the YRF to the CF. Integration with the 3-momentum part of the $\delta$-function imposes $p_a^{\rm YRF} = p_b^{\rm YRF} \equiv p^{\rm YRF}= \lambda(m_Y, m_a, m_b)/(2 m_Y)$, where $\lambda$ is the standard K\"all\'en function. Besides, the squared matrix element does not depend on the integration variables, it is just a number that can be directly rewritten as a function of the usual decay width, $\overline{|{\cal M}|}^2 = 8\pi m_Y^2 (\Gamma_{Y\to a,b}/p^{\rm YRF})$. Then, we can use the energy part of the delta function to obtain the expression
\begin{equation}
G_{Y\to a,b}(p_Y)= \frac{m_Y \Gamma_{Y\rightarrow a,b}}{2 E_Y} \int\limits_{-1}^{1} dc_{\theta}
\frac{e^{E_Y/T}}{(e^{E_a(c_\theta)/T} - \eta_a)(e^{E_b(c_\theta)/T} - \eta_b)}\ .
\label{eq:YwidthInMediumBeforecthetaint}
\end{equation}
This integral can also be evaluated analytically, yielding the final expression for the decay rate of a particle $Y$ of momentum $p_Y$ in the comoving frame into two particles $a$ and $b$, in the presence of the medium created by the latter
\begin{equation}
G_{Y\to a,b}(p_Y)=
\frac{m_Y \Gamma_{Y\rightarrow a,b}}{E_Y}
\frac{1 + \frac{m_Y T}{2 p_Y p^{\rm YRF}} 
\log\frac{\left( 1 - \eta_a e^{-E_a(1)/T} \right) \left( 1 - \eta_b e^{-E_b(-1)/T} \right)}
{\left( 1 - \eta_b e^{-E_b(1)/T} \right) \left( 1 - \eta_a e^{-E_a(-1)/T} \right)}}
{1 - \eta_a\eta_b e^{-E_Y/T}}~.
\label{eq:YwidthInMediumFinal}
\end{equation}
For future use, we also isolate the part of Eq.~\eqref{eq:YwidthInMediumFinal} that contains all the statistical mechanical information by defining the function $S$ through
\begin{equation}\label{eq:Sdefinition}
 G_{Y\to a,b}(p_Y) = 
\frac{m_Y \Gamma_{Y\rightarrow a,b}}{E_Y}
S\left( p_Y/T, x_Y, x_a, x_b, \eta_a, \eta_b \right)
\end{equation}
where $x_i \equiv m_i/T$.
\subsubsection{Back to the collision term}
Let us now turn back to the integrated collision term for the $Y \rightarrow a,b$ process. We can observe that Eq.~\eqref{eq:coll12dec} is similar to the one we just calculated, apart from the inclusion of an $f_Y$ factor in the integrand. It can be recast under the form
\begin{equation}
{\cal N}(Y\to a,b) = g_Y \left| \eta_Y \right| \int \frac{d^3 p_Y}{(2\pi)^3} \frac{e^{-E_Y/T}}{1-\eta_Y e^{-E_Y/T}}
\left[ 
\frac{m_Y \Gamma_{Y\rightarrow a,b}}{E_Y} S\left( \frac{p_Y}{T}, x_Y, x_a, x_b, \eta_a, \eta_b \right)
\right] \ .
\end{equation}
Next, we rewrite $d^3 p_Y = E_Y \sqrt{E_Y^2 - m_Y^2} dE_Y dc_{\theta} d\phi$. Integration over the angular variables yields a factor $4\pi$ and allows us to obtain the final expression for the integrated collision term \eqref{eq:coll12dec}
\begin{equation}
{\cal N}(Y\to a,b) =\frac{g_Y \left| \eta_Y \right|}{2\pi^2} m_Y^2 T \Gamma_{Y\rightarrow a,b} \tilde K_1 \left( x_Y, x_a, x_b, \eta_Y, \eta_a, \eta_b \right)
\label{coll1to2}
\end{equation}
where we have defined the function
\begin{align}\label{eq:K1tilde}
\tilde K_1 \left( x_1, x_2, x_3, \eta_1, \eta_2, \eta_3 \right) & \equiv
\frac{1}{(4\pi)^2 p_{\rm CM} T} \int \prod\limits_{i=1}^{3} \left( \frac{d^3 p_i}{E_i} \frac{1}{e^{E_i/T} - \eta_i} \right) e^{E_1/T} \delta^4 \left( P_1 - P_2 - P_3 \right) \\ \nonumber
& = x_1 \int\limits_{1}^{\infty}\frac{dz \sqrt{z^2 - 1} e^{-x_1 z}}{1 - \eta_1 e^{-x_1 z}} S\left( x_1\sqrt{z^2-1}, x_1, x_2, x_3, \eta_2, \eta_3 \right).
\end{align}
We are then left with a single integral to be evaluated numerically. Note that in the special case where $\eta_i=0$, the function $S$ in Eq.~(\ref{eq:Sdefinition}) is equal to unity and $\tilde K_1$ reduces to  the usual Bessel function of the second kind of degree one, $K_1 (m_Y/T)$.
\\
\\
In the special case of a mediator decaying into a pair of DM particles, $Y \rightarrow \chi\bar{\chi}$, we obtain
\begin{equation}
{\cal N}(Y\rightarrow \chi\bar{\chi}) =\frac{g_Y \left| \eta_Y \right|}{2\pi^2} m_Y^2 T \Gamma_{Y\rightarrow \chi\bar{\chi}} \tilde K_1 \left( x_Y, 0, 0, \eta_Y, 0, 0 \right).
\label{eq:coll1to2DM}
\end{equation}
\\
\\
Finally, by exploiting Eq.~\eqref{collsym}, we can also promptly obtain an expression for the collision term governing production of $Y$ from $a,b$ annihilations, namely
\begin{equation}
{\cal N}(a,b\to Y) = C_{ab} \frac{g_Y \left| \eta_a \eta_b \right|}{2\pi^2} m_Y^2 T \Gamma_{Y\rightarrow a,b} \tilde K_1 \left( x_Y, x_a, x_b, \eta_Y, \eta_a, \eta_b \right).
\label{eq:twotoone}
\end{equation}

We can now  proceed to the computation of the DM abundance. We will distinguish two cases, depending on whether the mediator is in thermal equilibrium with the SM or not.


\subsubsection{The case of a mediator in thermal equilibrium with the SM bath}
\label{sec:decay_thermal}

\begin{figure}[h]
  \centering
  \includegraphics[scale=0.75]{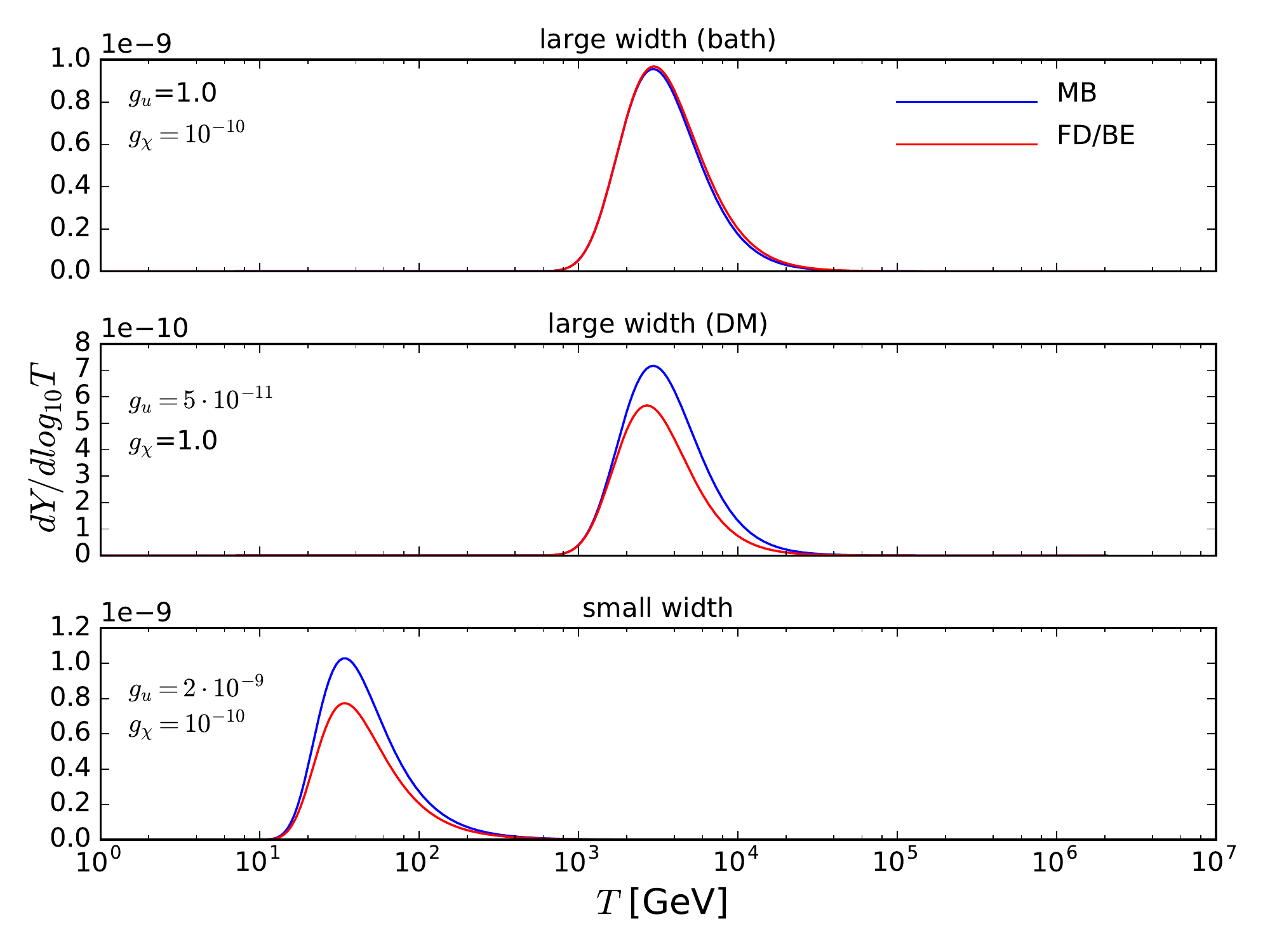} 
  \caption{Differential yield of DM production for a model of Dirac fermion dark matter with a $Z'$ mediator (\textit{cf} sec.~\ref{sec:vectorportal}), where the $Z'$ has a large width dominated by decays into SM particles (top), large width dominated by decays into DM pairs (middle) and very small width dominated by the decays into SM particles (bottom). In this figure we have taken $m_{Z'}=10^4$ GeV and considered only the process $u,\bar{u}\rightarrow \chi\bar\chi$. }
\label{fig:dYdT1}
\end{figure}

Consider, first, the case where the couplings of the mediator $Y$ to to ${\cal{B}}$-type particle pairs are large enough for it to be in equilibrium with the SM thermal bath, whereas its couplings (and, thus, the corresponding partial widths) involving DM particles are much smaller. Then, according to Eqs.~\eqref{yieldDM} and \eqref{coll1to2}, since $\eta_Y=\pm 1$ we can write: 
\begin{align}\label{yieldDMgeneral}
Y_\chi = 
\frac{g_Y}{2\pi^2} m_Y^2 
\Bigg[
\sum_X \bigg( & \Gamma_{Y\to \chi,X}
\int\limits_{T_0}^{T_R} \frac{dT}{\overline{H}(T) s(T)}  
\tilde K_1(x_Y,0,x_X,\eta_Y,0,\eta_X) \bigg) \\ \nonumber
+
2 & \Gamma_{Y\to \chi\bar{\chi}}
\int\limits_{T_0}^{T_R} \frac{dT}{\overline{H}(T) s(T)}  
\tilde K_1(x_Y,0,0,\eta_Y,0,0)
\Bigg] \ .
\end{align}
Although in \micro~the yield is obtained by evaluating Eq.~\eqref{yieldDMgeneral}, it is instructive to also present an approximate expression for the -- often occuring -- scenario of a mediator decaying into a pair of DM particles. In this case, the last expression can be recast as
\begin{equation}
Y_\chi = \dfrac{g_Y}{m_Y^2}\Gamma_{Y\to\chi\bar\chi} \int\limits_{m_Y/T_R}^{m_Y/T_0} x_Y^2 e^{-x_Y}dx_Y
\left[\dfrac{x_Y e^{x_Y} \tilde K_1(x_Y,0,0,\eta_Y,0,0)}{\pi^2 \overline{H}(T) s(T)T^{-5}}\right]~.
\label{yieldDMa}
\end{equation}
The term in square brackets, which we will denote as ${\cal P}(x_Y,\eta_Y)$, has a mild dependence on the temperature $T$, while the quantity $ x_Y^2e^{-x_Y}dx_Y$ peaks at $x_Y= 3$ with a width $\sigma\approx 1.7$. Then, for large enough $T_R$ we can approximate the DM production yield as:
\beq
Y_\chi = \dfrac{g_Y \Gamma_{Y\to\chi\bar\chi}}{m_Y^2}\int\limits_0^\infty dx_Y~x_Y^2e^{-x_Y}{\cal P}(x_Y,\eta_Y) \approx
2\dfrac{g_Y \Gamma_{Y\to\chi\bar\chi}}{m_Y^2}{\cal P}(3,\eta_Y)  ~.
\eeq
Performing the integration we get (for a bosonic mediator)
\beq
Y_\chi \approx 0.685\, \dfrac{g_Y M_{Pl} \Gamma_{Y\to\chi\bar\chi}}{m^2_Y}\left(\dfrac{1+\frac{1}{3}\frac{dh_{\rm eff}(T)}{dT}}{\sqrt{g_{\rm eff}(T)}h_{\rm eff}(T)}\right)_{T=m_Y/3}~.
\eeq
This approximate result, a similar form of which was also obtained in \cite{Hall:2009bx}, works with a precision of $\approx 1\%$ for mediator masses $m_Y>3$ GeV. Moreover, when $T_R>2m_Y$ the dependence on the reheating temperature is below the 1\% level. 
\\
\\
Finally, since throughout our analysis we consistently use the full Bose-Einstein/Fermi-Dirac distribution functions for initial and final state bath particles, we can also quantify the accuracy of the commonly adopted Maxwell-Boltzmann approximation. Neglecting the dark matter distribution function, statistics enter either through the mediator or through the final state particle that is produced in association with DM. At the level of the former, we find that if the mediator is a boson, using a Bose-Einstein distribution instead of a Maxwell-Boltzmann one leads to a 3.5\% increase  in the DM abundance, see for example Fig.~\ref{fig:dYdT1}-top panel. This figure also illustrates that the dominant contribution occurs near $m_Y/3$. For a fermion mediator, the use of Fermi-Dirac statistics leads instead to a 2.5\% decrease in the DM abundance. Moreover, in this case the final state necessarily involves two different particles. If one of them is a SM fermion, then the statistics factor can be much larger especially when there is a small mass splitting between the mediator and the DM particle, as will be seen in section~\ref{sec:t-channel}.

\subsubsection{The case of a  mediator not in chemical equilibrium with the SM bath}
\label{sec:noteq}

For sufficiently suppressed couplings between the mediator and the bath particles, the two sectors will never reach chemical equilibrium. A Boltzmann equation for the mediator should, hence, be considered together with the one corresponding to DM. In this regime the decay width of the mediator is so small that it decays at a temperature $T\ll m_Y$ and, so, whether or not \textit{kinetic} equilibrium between the mediator and the bath particles is attained is fairly irrelevant for the ensuing DM abundance. In order to justify this statement, below we present the evolution of the mediator's abundance in two opposite regimes: 1) assuming that it is in kinetic equilibrium with the bath, which means that its temperature coincides with the SM one, and 2) assuming that it is kinetically decoupled from the SM bath. The DM abundance value obtained through the two methods coincides at the percent level.
\\
\\
{\bf Mediator in kinetic equilibrium with the SM bath}
\\
\\
Neglecting mediator production from FIMP annihilations, its evolution is described by the Boltzmann equation:
\begin{equation}
\dot{n}_Y + 3Hn_Y = {\cal{N}}_{{\rm bath} \rightarrow Y} - {\cal{N}}_{Y \rightarrow {\rm all}}~.
\end{equation}
We can, instead, trade the number density $n_Y$ for the abundance $Y_Y$ and, using Eqs.~\eqref{eq:coll1to2DM} and \eqref{eq:twotoone}, write
\begin{equation}
\frac{dY_Y}{dt} = \frac{g_Y m_Y^2 T}{2 \pi^2 s(T)} 
\sum_{a,b} \left[ \Gamma_{Y\rightarrow a,b} \left( \left|\eta_a \eta_b\right| - \left|\eta_Y\right| \right) \tilde{K}_1 (x_Y, x_a, x_b, \eta_Y, \eta_a, \eta_b) \right]~,
\label{eq:kineqevolY}
\end{equation}
where $\eta_i = 0$ for FIMPs and $=\pm 1$ for bath particles. The mediator's yield $Y_Y$ and parameter $\eta_Y$ are related through Eqs.~\eqref{eq:numberenergydensitydef} and ~\eqref{eq:FD-BE} as
\beq 
Y_Y = \dfrac{g_Y |\eta_Y| T m_Y^2 } {2\pi^2 s(T)} \tilde K_2(x_Y,\eta_Y)~, 
\label{eq:eta_med}
\eeq
where
\beq
\tilde K_2(x,\eta) = \dfrac{1}{x^2}\int\limits_0^\infty \dfrac{y^2 dy}{e^{\sqrt{x^2+y^2}}-\eta}~,
\eeq
which reduces to  the Bessel function of second order and degree two, when Maxwell-Boltzmann statistics (corresponding to $\eta=0$) is assumed, \textit{i.e.}  $\tilde K_2(x,0) = K_2(x)$.
The evolution equation for the mediator, Eq.~\eqref{eq:kineqevolY}, can be rewritten in a more intuitive way by defining an average width ${\Gamma}^{\rm bath}_{Y\rightarrow a,b}$ for $Y$ in the presence of the thermal bath, along with a boost factor $\gamma(x_Y, \eta_Y)$ as
\begin{align}
\label{eq:avwidth}
{\Gamma}^{\rm bath}_{Y\rightarrow a,b}  & \equiv \frac{\tilde{K}_1 (x_Y, x_a, x_b, \eta_Y, \eta_a, \eta_b)}{\tilde{K}_1 (x_Y, 0, 0, \eta_Y, 0, 0)} \Gamma_{Y\rightarrow a,b} \\
\gamma(x_Y, \eta_Y) & \equiv \dfrac{\tilde K_2(x_Y,\eta_Y)}{\tilde K_1(x_Y,0,0,\eta_Y,0,0)}~.
\label{eq:avboost}
\end{align}
Note that the normalisation $\tilde K_1(x,0,0,\eta,0,0)^{-1}$ in Eqs.~\eqref{eq:avwidth} and \eqref{eq:avboost} is simply conventional.
\\
\\
Using these definitions, we can indeed rewrite Eq.~\eqref{eq:kineqevolY} as
\begin{equation}
\dfrac{dY_Y}{dt} = -\dfrac{1}{\gamma(x_Y,\eta_Y)} ({\Gamma}^{\rm bath}_{Y\to {\rm all}} Y_Y - {\Gamma}^{\rm bath}_{Y\to {\rm bath}} \bar Y_Y)~,
\label{eq:Boltz_med}
\end{equation}
where
\begin{equation}
\bar Y_Y = \bar Y_Y(T,m_Y,\eta_Y) = \dfrac{g_Y}{2\pi^2}\dfrac{Tm_Y^2}{s(T)}\tilde K_2(x_Y,\eta_Y)~.
\label{eq:med_eq}
\end{equation}
\\
\\ 
Some comments are in order at this point:
\begin{itemize}
\item The production term (second term) in Eq.~\eqref{eq:Boltz_med} is mathematically equal to the sum of all the 2$\to$1 processes where two bath particles create one mediator if they were in equilibrium. Of course the mediator (by assumption in this section) is not in equilibrium: this is just a mathematical trick to render the computations easier.
\item The expression \eqref{eq:Boltz_med} depends on the chemical potential of the mediator through the parameter $\eta_Y$. For mediators in chemical equilibrium $\eta_Y=1$. However, if they are not in equilibrium, $\eta_Y$ has to be calculated. This is done by finding an  iterative solution to  Eq.~\eqref{eq:eta_med} in the form  $\eta_Y=\eta_Y(Y_Y)$ and substituting it in Eq.~\eqref{eq:med_eq}.
Note that $\eta_Y$ can be much greater than one (in absolute value) if the mediator's population is much larger than its equilibrium population. 
\end{itemize}
Finally, once we know the mediator yield $Y_Y$ from equations \eqref{eq:Boltz_med} and \eqref{eq:eta_med}, the DM abundance can be directly computed using
\beq
\frac{dY_\chi}{dt}= 
\frac{g_Y m_Y^2 T |\eta_Y|}{2 \pi^2 s(T)}  \left[2  \Gamma_{Y\rightarrow \chi\chi} \tilde{K}_1 (x_Y, 0, 0, \eta_Y, 0,0) 
+\sum\limits_{a\in {\cal B}}  \Gamma_{Y\rightarrow a\chi} \tilde{K}_1 (x_Y, x_a, 0, \eta_Y, \eta_a, 0) \right]
\eeq
which can be simply rewritten as
\beq
\frac{dY_{\chi}}{dt}= \frac{Y_Y}{\gamma(x_Y,\eta_Y)} \left( \Gamma^{\rm bath}_{Y\rightarrow \chi X}+ 2 \Gamma^{\rm bath}_{Y\rightarrow \chi\chi}\right)
\eeq
\\
So far we have assumed that the mediator has the same temperature $T$ as the bath. This assumption is not necessarily justified, since we are analysing the situation where the mediator is not in chemical equilibrium with the SM. In order to check how this assumption affects the estimated dark matter abundance, we now study the evolution of the mediator in the opposite regime of complete kinetic decoupling.
\\
\\
{\bf Kinetically decoupled mediator}
\\
\\
Our starting point in order to track the evolution of the mediator phase-space distribution $f_Y(p_Y,t)$, is the \textit{un}-integrated form of the Boltzmann equation. For fixed mediator energy $E_Y$, the latter reads \cite{Kolb:1990vq}:
\begin{align}\label{eq:BoltzmannUnintegrated}
E_Y \left( \dot{f_Y} - H p_Y \frac{\partial f_Y}{\partial p_Y} \right) & = 
\frac{1}{2} \int \frac{d^3 p_a}{(2\pi)^3 (2 E_a)} \frac{d^3 p_b}{(2\pi)^3 (2 E_b)} (2\pi)^4 \delta^4 \left(P_Y - P_a - P_b\right) \\ \nonumber
& \bigg[ C_{ab} f_a f_b (1\mp f_Y) \left|{\cal{M}}_{a,b \to Y} \right|^2 - 
f_Y (1\mp f_a) (1\mp f_b) \left|{\cal{M}}_{Y \to a,b} \right|^2 \bigg]~,
\end{align}
where the kinetic decoupling assumption is reflected by the fact that we have ignored terms amounting to energy redistribution. The depletion (second) term is very similar to the expression $G_{Y\to a,b}$ that we computed in Section \ref{sec:onetotwo} in terms of the $S\left( p/T, x_Y, x_a, x_b, \eta_a, \eta_b \right)$ function, Eq.~\eqref{eq:Sdefinition}. The first term, on the other hand, can be manipulated in a similar manner as in Eq.~\eqref{collsym}. Then, Eq.~\eqref{eq:BoltzmannUnintegrated} can be rewritten as
\begin{align}\label{eq:BoltzFY}
\dot f_Y & - Hp_Y \frac{\partial f_Y(p_Y)}{\partial p_Y} = 
\dfrac{m_Y}{E_Y}\sum_{a,b} S\left(p_Y/T,x_Y,x_a,x_b,\eta_a,\eta_b\right) \\ \nonumber
& \times\bigg[ |\eta_a\eta_b|e^{-E_Y/T} \left(1\mp f_Y(p_Y)\right)-f_Y(p_Y) \bigg] \Gamma_{Y\to a,b}~.
\end{align}
To solve this partial differential equation we use the method of characteristics. We first reduce it to a set of coupled ordinary differential equations by introducing the new variables
\begin{eqnarray}
\phi(\pi,T) &=&f_Y(p(\pi,T))\\
p(\pi,T)&=&\pi\sqrt[3]{\frac{s(T)}{s(T_0)}}
\end{eqnarray}
where $p(\pi,T)$ represents the momentum of a particle at a temperature $T$, which is reduced to $\pi$ at a temperature $T_0$ due to  the expansion of the Universe. Then, Eq.~\eqref{eq:BoltzmannUnintegrated} amounts to an evolution equation for $\phi$ which does not contain any  partial derivatives, once we switch from a time to a temperature derivative using Eq.~\eqref{eq:timetemperaturerelation}. Namely, we obtain
\begin{eqnarray}
\nonumber
   \frac{d\phi(\pi,T)}{dT}&&= -\frac{m_Y}{E \overline{H}(T)T} \times\\
\label{eq:phi_pi_eq}
 && \sum \limits_{a,b}  \Gamma_{Y\to a,b} S(\frac{p(\pi,T)}{T}, \mu_Y,\mu_a,\mu_b,\eta_a,\eta_b)
 \left(|\eta_a\eta_b| e^{-\frac{E}{T}} (1 \pm \phi(\pi))-\phi(\pi)\right) 
\end{eqnarray} 
where $E=\sqrt{p(\pi,T)^2+m_Y^2}$. We also
 introduce the DM distribution function $\phi_\chi(\pi,T)$ which represents the number of DM particles produced  from the decay of a  mediator with momentum $p(\pi,T)$. Its evolution equation reads in turn
\begin{eqnarray}
\nonumber
  \frac{d{\phi_\chi(\pi,T)}}{dT} &=&  -\frac{m_Y g_Y \phi(\pi,T)}{E \overline{H}(T)T} \\
\label{eq:n_pi_eq}
      && \left( 2 \sum \limits_{a,b \in \tilde{\cal{F}}}  \Gamma_{Y\to a,b} +
             \sum \limits_{a \in \tilde{\cal{B}} ,b \in \tilde{\cal{F}}}  \Gamma_{Y\to a,b} S(\frac{p(\pi,T)}{T}, x_Y,x_a,x_b,\eta_a,0)\right)~.
\end{eqnarray}

After integrating Eqs.~\eqref{eq:phi_pi_eq} and \eqref{eq:n_pi_eq} over $T\in [T_0,T_R]$, the number density of DM particles at temperature $T_0$  can be calculated from
\begin{equation}
\label{N_sterile}
         n_{\chi}= \frac{1}{(2\pi)^3}\int  \phi_\chi(\pi) d^3\pi~.
\end{equation} 
In practice, in the code we choose a grid  for $\pi$, solve the two coupled differential equations \eqref{eq:phi_pi_eq} and \eqref{eq:n_pi_eq} for each element of the grid, interpolate the result and calculate the integral Eq.~\eqref{N_sterile}.
 
We have checked that the result coincides at the percent level with the one obtained assuming the mediator is in kinetic equilibrium with the SM bath. For both methods we have also checked explicitly the impact of the statistics factor both for the case where a bosonic mediator decays predominantly into DM particles and into SM ones. For a mediator coupling mostly to fermions we find that the Fermi-Dirac statistics leads to a decrease of about 25\% in the final DM abundance regardless of whether the mediator decays rapidly into DM (see Fig.~\ref{fig:dYdT1} middle panel), or its decay is delayed (top panel). If, on the other hand, the mediator couples mostly to bosons the statistics factor leads to an increase in the DM abundance by a factor of up to 80\%. Explicit examples for this scenario will be presented in Section~\ref{sec:models}.

\subsection{$2\to 2$ processes}
\label{sec:2to2}
Having presented our approach for DM production through mediator decays, we next consider  $1,2\to a,b$ processes where $1,2$ are bath particles and either or both of $a$ and $b$ are FIMPs. In this case the integrated collision term reads:
\begin{eqnarray}\label{eq:22}
{\cal N}(1,2\to a,b) &=& C_{12} \int \dfrac{d^3p_1}{(2\pi)^32E_1} \dfrac{d^3p_2}{(2\pi)^32E_2} \dfrac{d^3p_a}{(2\pi)^32E_a} \dfrac{d^3p_b}{(2\pi)^32E_b}\\ \nonumber
&\times& (2\pi)^4\delta^4(P_1+P_2-P_a-P_b)|{\cal M}|^2 f_1 f_2 (1\mp f_a)(1\mp f_b)~.
\end{eqnarray}
If both $a$ and $b$ are FIMPs, we can approximate $(1\mp f_a)(1\mp f_b) \simeq 1$ and this expression is very similar to the one obtained in the case of standard thermal freeze-out, \textit{cf e.g.} \cite{Gondolo:1990dk}. It can be manipulated in a similar manner and rewritten in terms of the $\tilde K_1$ function defined in Eq.~\eqref{eq:K1tilde} as
\begin{equation}
{\cal N}(1,2\to \chi_a,\chi_b) = \frac{T g_1g_2 \left|\eta_1 \eta_2 \right|}{8\pi^4} C_{12}\int ds~(p^{\rm CM}_{1,2})^2\sqrt{s}\sigma(s) \tilde K_1(\sqrt{s}/T,x_1,x_2,0,\eta_1,\eta_2)~.
\label{coll2to2} 
\end{equation} 
In the case where the $2\to2$ reaction amounts to the production of a single feebly coupled particle along with a bath particle $a\equiv X$, the situation becomes more involved. In this case one has to actually perform the full 12-dimensional integration of Eq.~\eqref{eq:22}, after replacing $f_b\to 0$. Given that performing multiple integrations is computer-time consuming, in \micro~we instead use an empirical approximation that entails introducing a correction factor in the integrand of Eq.~\eqref{coll2to2} as
\begin{equation}
\frac{\tilde K_1(\sqrt{s}/T,x_\chi,x_X,0,0,\eta_X)}{\tilde K_1(\sqrt{s}/T,x_\chi,x_X,0,0,0)} \ ,
\label{corrcoll2to2} 
\end{equation}
where the denominator is simply $K_1(\sqrt{s}/T)$. We have compared this approximation to the full numerical calculation of Eq.~\eqref{eq:22} for the model discussed  in section~\ref{sec:t-channel} and have found that this simple correction factor reproduces the exact result with a precision of 2\%. Here we assume that the $X$ particle thermalises rapidly.

\subsubsection{The special case of an $s$-channel resonance}\label{sec:sresonance}

When summing over all DM production channels (\textit{cf} Section \ref{sec:routines}), \micro~will compute the abundance by adding the yield from all $2\rightarrow 2$ processes for any point in parameter space. When DM production is dominated by the decay of the mediator, we have to ensure that the result from each $2\rightarrow 2$ process matches the one obtained from mediator decays. To this end we modify, for each temperature, the total width of the mediator appearing in the relevant cross sections\footnote{This discussion focuses on dark matter pair-production processes or, eventually, processes involving only FIMPs in the final state. The case of $\chi X$, $X \in {\cal{B}}$ final states will be treated in a future release of \micro.}.

In the narrow-width approximation (NWA), the Breit-Wigner cross section for DM pair-production via the process $1,2\to Y\to \chi\bar \chi$ in the presence of a SM bath reads:
\begin{equation}
\sigma(s) = \dfrac{g_Y}{g_1 g_2}\dfrac{1}{C_{12}} \dfrac{4\pi^2 m_Y}{(p^{\rm CM}_{1,2})^2}  \dfrac{\Gamma_{Y\to 1,2} \Gamma_{Y\to\chi\bar\chi}}{\Gamma_{\rm tot}}\delta(s-m_Y^2)~,
\end{equation}
where we have used the expressions for the partial widths to parametrize the dependence on the couplings, thus it is the partial width in vacuum $\Gamma_{Y\to 1,2}$
that appears in the numerator. Substituting this expression in Eq.~\eqref{coll2to2} allows us to write:
\begin{equation}
\label{eq:n12}
{\cal N}(1,2\to Y \to \chi\bar\chi) = \dfrac{T g_Y \left|\eta_1 \eta_2 \right|}{2\pi^2} m_Y^2 
\dfrac{\Gamma_{Y\to 1,2} \Gamma_{Y\to \chi\bar\chi}}{\Gamma_{\rm tot}} \tilde K_1(x_Y,x_1,x_2,0,\eta_1,\eta_2)~.
\end{equation}
For this equation to match the one used for the decay of the mediator, Eq.~\eqref{eq:coll1to2DM}, the total width of the latter is corrected by two temperature-dependent factors. The first one simply takes into account the chemical potential of the mediator
\begin{equation}
\Gamma_{\rm tot} \rightarrow  \Gamma_{\rm tot} \frac{\tilde K_1(x_Y,x_1,x_2,0,\eta_1,\eta_2)}{\tilde K_1(x_Y,x_1,x_2,\eta_Y,\eta_1,\eta_2)}.
\end{equation} 
\\
\\
The second factor is required to replace the branching ratio $BR_{Y\to \chi\chi}=\Gamma_{Y\to \chi\bar\chi}/\Gamma_{\rm tot}$ in Eq.~\eqref{eq:n12} by the corresponding one in the thermal bath. Note that we cannot simply correct Eq.~\eqref{coll2to2} by an overall factor, since doing so would also modify the off-resonance contribution. For this reason, we rescale the total width for each temperature. 

If the mediator's decay width is small, its decay is not instantaneous but rather happens at a temperature $T_d$ smaller than its production temperature. The dependence on $T_d$ lies inside $\Gamma^{\rm bath}_{\rm tot}$, so we replace the mediator branching ratio into DM using the following ansatz:
\begin{equation}\label{eq:effBR}
BR_{Y\to \chi\chi} \to BR^{\rm eff}_{Y\to \chi\chi}(T_c) = \int\limits_0^1 \frac{\Gamma^{\rm bath}_{Y\to \chi\chi}(T_d(P))}{\Gamma^{\rm bath}_{\rm tot}(T_d(P))} dP
\end{equation}
where $P$ is the probability of the mediator created at temperature $T_c$ to survive until a temperature $T_d$. An integral representation for $P(T_d)$ can be obtained through Eq.~\eqref{eq:Boltz_med}. It reads
\beq
P(T_d) = e^{-L(T_d)},~~~~L(T_d) = \int\limits_{T_d}^T \dfrac{\Gamma^{\rm bath}_{\rm tot}(T')}{\gamma(T')\overline{H}(T')T'}dT'~, 	
\eeq
where $\overline{H}(T)$ was defined in Eq.~\eqref{eq:Hprimedef} and $\gamma(T')$ in Eq.~\eqref{eq:avboost}.
\\
\\
In practice, before the calculation of the contribution of each $2\to 2$ reaction to DM production, \micro~detects all $s$-channel resonances in the corresponding cross section. Using the formalism presented in Section \ref{sec:noteq}, and assuming kinetic equilibrium of the mediator with the SM thermal bath, the functions $\eta_Y(T)$, $\Gamma^{\rm bath}_{Y\to \chi\chi}(T)$ and $\Gamma^{\rm bath}_{\rm tot}(T)$ are tabulated and the effective width that can be extracted from Eq.~\eqref{eq:effBR} is computed. Finally, this width is substituted in the matrix element expression of the relevant $2\to 2$ reaction.

\section{Functions of \micro~}
\label{sec:micro}

Having presented all the necessary formalism for freeze-in dark matter production and the way it is implemented in \micro, we now describe some practical aspects of the code and point out some limitations that will be dealt with in future distributions.

\subsection{Constants and auxiliary functions}
All physical constants used in calculations of freeze-out or freeze-in scenarios are defined  in the  file 
\verb|include/micromegas_aux.h|. The ones  used explicitly in  freeze-in scenarios  are listed in Table~\ref{tab:constants}. Some auxiliary functions that can be called anywhere in the code are also provided:

\begin{table}[h]
\begin{tabular}{llll}
\hline
Name & Value & Units & Description\\\hline
\verb|MPlank| &     $1.22091\times 10^{19}$  &  GeV & Planck mass    \\
\verb|EntropyNow| &  $2.8912\times 10^{9}$    & ${\rm m}^{-3}$ & Present day entropy, $s_0$     \\
\verb|RhoCrit100| & 10.537     & ${\rm GeV m}^{-3}$  & $\rho_c/h^2$  or $\rho$ for $H=100$ km/s/Mpc \\\hline
\end{tabular}
\caption{Some useful constants included in \micro.}
\label{tab:constants}
\end{table}


\begin{itemize}
\item \verb|gEff(T)|  - returns the effective number of degrees of freedom for the energy density of radiation at a bath temperature \verb|T| (\textit{cf} Eq.~\eqref{geff}),   only SM particles are included.
\item \verb|hEff(T)|  - returns the effective number of degrees of freedom for the entropy density of radiation at a bath temperature \verb|T| (\textit{cf} Eq.~\eqref{heff}).
\item \verb|Hubble(T)| - returns the Hubble expansion rate at a bath temperature \verb|T|. This applies to the radiation-dominated era and is valid for \verb|T| $\gtrsim 100$ eV.
\item \verb|hEffLnDiff(T)| - returns the derivative of $h_{eff}$ with respect to the bath temperature, $\frac{d\log(h_{eff}(T))}{d\log(T)}$.
\item \verb|Stat2(|$p,x_Y,x_1,x_2,\eta_1,\eta_2$\verb|)|, returns the $S$ function defined through Eqs.~\eqref{eq:YwidthInMediumFinal} and \eqref{eq:Sdefinition}, that takes into account particle statistical distributions for the decay of a mediator $Y$ of fixed momentum $p$. 
\item \verb|K1to2(|$x_1,x_2,x_3,\eta_1,\eta_2,\eta_3$\verb|)|, returns the $\tilde{K}_1$ function defined in Eq.~\eqref{eq:K1tilde}, that takes into account particle statistical distributions.   
\end{itemize}
\subsection{Freeze-in routines}
\label{sec:routines} 

Several routines are provided in \micro~to compute the DM abundance in freeze-in scenarios. These can be found in the file \verb|sources/freezein.c|. The first line of this file contains the statement
\begin{center}
\verb|//#define NOSTATISTICS|
\end{center}
This statement can be uncommented for \micro~ to compute the relic density assuming a Maxwell-Boltzman distribution. This option is faster. 
\\
\\
We remind the user that the code does \textit{not} check whether or not a particle is in thermal equilibrium with the SM thermal bath and that it is the responsibility of the user to specify which particles belong to the bath, ${\cal B}$, or are out of equilibrium, ${\cal F}$. This can be done through the function\\
\noindent
$\bullet$ \verb|toFeebleList(particle_name)|\\
which assigns the particle \verb|particle_name| to the list of feebly interacting ones (\textit{i.e.} those which belong to ${\cal F}$). Feebly interacting particles can be odd or even.   
This function can be called several times to include more than one particle.  All odd or even particles that are not in this list are assumed to be in thermal equilibrium with the SM and belong to ${\cal B}$. 
The treatment of the particles that belong to  ${\cal F}$ for the computation of $\Omega h^2$ within the freeze-in routines is described below. 
Calling \verb|toFeebleList(NULL)| will reassign all particles to ${\cal B}$.
\\
\\
\\
The actual computation of the freeze-in dark matter abundance can be performed with the help of three functions:
\\
\\
$\bullet$ \verb|darkOmegaFiDecay(TR, Name, KE, plot) | \\
calculates the DM abundance from the  decay of the particle \verb|Name| into all odd FP. \verb|TR| is the  reheating temperature and \verb|KE| is a switch to specify whether the decaying particle is in kinetic equilibrium (\verb|KE|=1) or not (\verb|KE|=0) with the SM.  If the decaying particle has not been declared as feeble (see above) then Eq.~\eqref{yieldDMgeneral} is solved. Otherwise the methods described in Section \ref{sec:noteq} are used, specifically Eqs.~\eqref{eq:eta_med},\eqref{eq:med_eq},\eqref{coll1to2} and \eqref{yieldDM} when \verb|KE=1| and   Eqs.~\eqref{eq:BoltzFY},\eqref{N_sterile} when \verb|KE=0|. Numerically, the latter two methods give very similar results, however the \verb|KE=1| option is faster. The switch \verb|plot=1| displays on the screen  $Y(T)$ for the decaying particle and $dY/d\log(T)$ for DM.

This routine operates under a set of underlying assumptions and limitations. First, as already discussed in Section \ref{sec:generalFI}, we assume that all odd FPs will decay into the lightest one, which is the DM. Consequently, if at least one final state contains dark sector FIMPs that decay predominantly into bath particles, $\verb|darkOmegaFiDecay|$ will overestimate the corresponding DM yield. Simply put, from a DM yield standpoint in \micro~each dark sector FIMP is equivalent to a dark matter particle. Secondly, if particle $\verb|Name|$ can decay into a final state involving one or two mediators which can subsequently decay into DM pairs, the contribution of the latter decay is not taken into account. Only $2$-body final states are currently supported.
\\
\\
$\bullet$ \verb|darkOmegaFi22(TR, Process, vegas, plot, &err)|  \\
calculates the  DM abundance taking into account only DM production via the  $2\to2$ process defined by the parameter {\tt Process}. For example \verb|"b,B ->~x1,~x1"|  for the production of DM (here \verb|~x1|) via $b\bar{b}$ scattering. This routine allows the user to extract the contribution of individual processes. \verb|TR| is the  reheating temperature. When  the switch \verb|vegas=1|, the collision term is integrated directly using Eq.~\eqref{eq:22}. The execution time for this option is quite long,  it is  intended mostly for precision checks. The switch \verb|plot=1| displays on the screen  $dY/d\log(T)$ for DM. Note that the
temperature profile for DM production obtained by \verb|darkOmegaFiDecay| and \verb|darkOmegaFi22| can be different. For example, in the case of an $s$-channel resonance, the temperature for DM production corresponds to the one of the mediator decay for  \verb|darkOmegaFiDecay|  and  the temperature at which the mediator is created for   \verb|darkOmegaFi22|.
Finally, {\tt err} is the returned error code, which has the following meaning\\
1: the requested processes does not exist\\
2: $2\rightarrow 2$ process is expected\\
3: cannot calculate local parameters  / some constrained parameters cannot be calculated\\
4: the reheating temperature is too small ($T_R<1 {\rm keV}$)\\
5: one of the incoming particles belongs to ${\cal F}$\\
6: None of the outgoing particles are odd  and feeble\\
7: The integral over temperature does not converge\\
8: The integral over energy does not converge\\
9: The angular integral does not converge\\
10: There is an  on-shell particle in t/u channel\\
11: Cannot treat the small angle pole contribution.\\
12: Lost of precision due to diagram cancellation.\\
When substituting \verb|NULL| for the error code, the error message is displayed on the screen.
\\
\\
$\bullet$ \verb|darkOmegaFi(TR, &err)|  \\
calculates the  DM abundance after summing over all $2\to2$ processes involving  particles in the bath ${\cal B}$ in the initial state  and at least one odd particle in ${\cal F}$  in the final state, following the method described in Section~\ref{sec:2to2}. The routine checks the decay modes of all bath particles and if one of them has no decay modes into two other bath particles, the $2\to2$ processes involving this particle are removed from the summation and instead the contribution to the DM abundance computed from the routine \verb|darkOmegaFiDecay| is included in the sum.  This is done to avoid appearance of poles in the corresponding $2\to2$ cross-section. We recommend for such models to compute individual $2\to2$ processes with \verb|darkOmegaFi22| described above. As before, we assume that all odd FIMPs will decay into the lightest one which is the DM. {\tt err} is the returned error code, \verb|err=1| if feeble particles have not been defined. 

Note that in the case of models with two distinct dark sectors containing two dark matter candidates, the odd FIMPs of each dark sector are assumed to decay into the lightest particle (the DM) of the same class. The function \verb|darkOmegaFi(TR, &err)| computes the total DM abundance and the two contributions can be disentangled by calling the function \verb|sort2FiDm( &omg1,&omg2)|.
\\
\\
\\
Finally, \\
\noindent
$\bullet$ \verb|printChannelsFi(cut,prcnt,filename)|  \\
writes into the file \verb|filename| the contribution of different channels to $\Omega h^2$. The \verb|cut| parameter specifies the lowest relative contribution to be printed.   If
\verb|prcnt| $\neq 0$, the contributions are given in percent.   
The routine \verb|darkOmegaFi| fills the array \verb|omegaFiCh| which contains the contribution of different channels ($2\rightarrow 2$ or $1\rightarrow 2$) to $\Omega h^2$.
\verb|omegaFiCh[i].weight| specifies the relative weight of the i\verb|t|-h channel, whereas \verb|omegaFiCh[i].prtcl[j]| (with \verb|j|=0,...,4) defines the particle names for the \verb|i|-th channel.
The last record in the array  \verb|omegaFiCh| has zero weight and \verb|NULL| particle names.
\\
\\
\\
Note that if no particle has been declared as being feebly interacting, the freeze-out routines \verb|darkOmega|, \verb|darkOmegaFO|, and \verb|darkOmega2| ~\cite{Belanger:2014vza} will work exactly like in previous versions of \micro. A non-empty list of FIMPs, however, will affect these routines since \micro~will exclude all odd particles in this list from the computation of the relic density via freeze-out. For example, if the DM is a bath particle, excluding FPs can impact the freeze-out computation of $\Omega h^2$ when they are nearly degenerate in mass with the DM, hence could potentially contribute to coannihilation processes. Note also that if, for example, the lightest odd particle (the DM) belongs to ${\cal F}$ and the user computes the freeze-out abundance of the lightest odd bath particle, the resulting  value of $\Omega_{LBP} h^2$  will be automatically rescaled by a factor $M_{LFP}/M_{LBP}$, where $M_{LBP}$($M_{LFP}$) is the mass of the lightest odd particle in ${\cal B}$ (${\cal F}$). Conversely, if the DM belongs to ${\cal B}$ and the user computes the freeze-in abundance of the lightest feeble odd particle, the corresponding result for $\Omega_{LFP} h^2$ will be rescaled by $M_{LBP}/M_{LFP}$. In other words, the answer obtained in both of these cases corresponds to the predicted density of the dark matter particles and not the heavier ones. Besides, \micro~ does not check whether the decay rate of an odd particle to the feeble particles is much smaller than $H(T_{FO})$ which would justify the fact that they should not be included in the freeze-out computation.

\section{Sample models and numerical results} 
\label{sec:models}

We now move on to present some numerical results obtained with \micro~for a few simple dark matter models. These models are included by default in the public distribution of \micro$5.0$~and can serve as guidelines for the implementation of additional New Physics scenarios. Moreover, we will give special emphasis to the impact of the statistical distribution of bath particles, an effect which has only rarely been considered in the literature. 

\subsection{Vector portal}\label{sec:vectorportal}

\begin{figure}
 \begin{center}
 \subfloat[]{\includegraphics[width=0.3\textwidth]{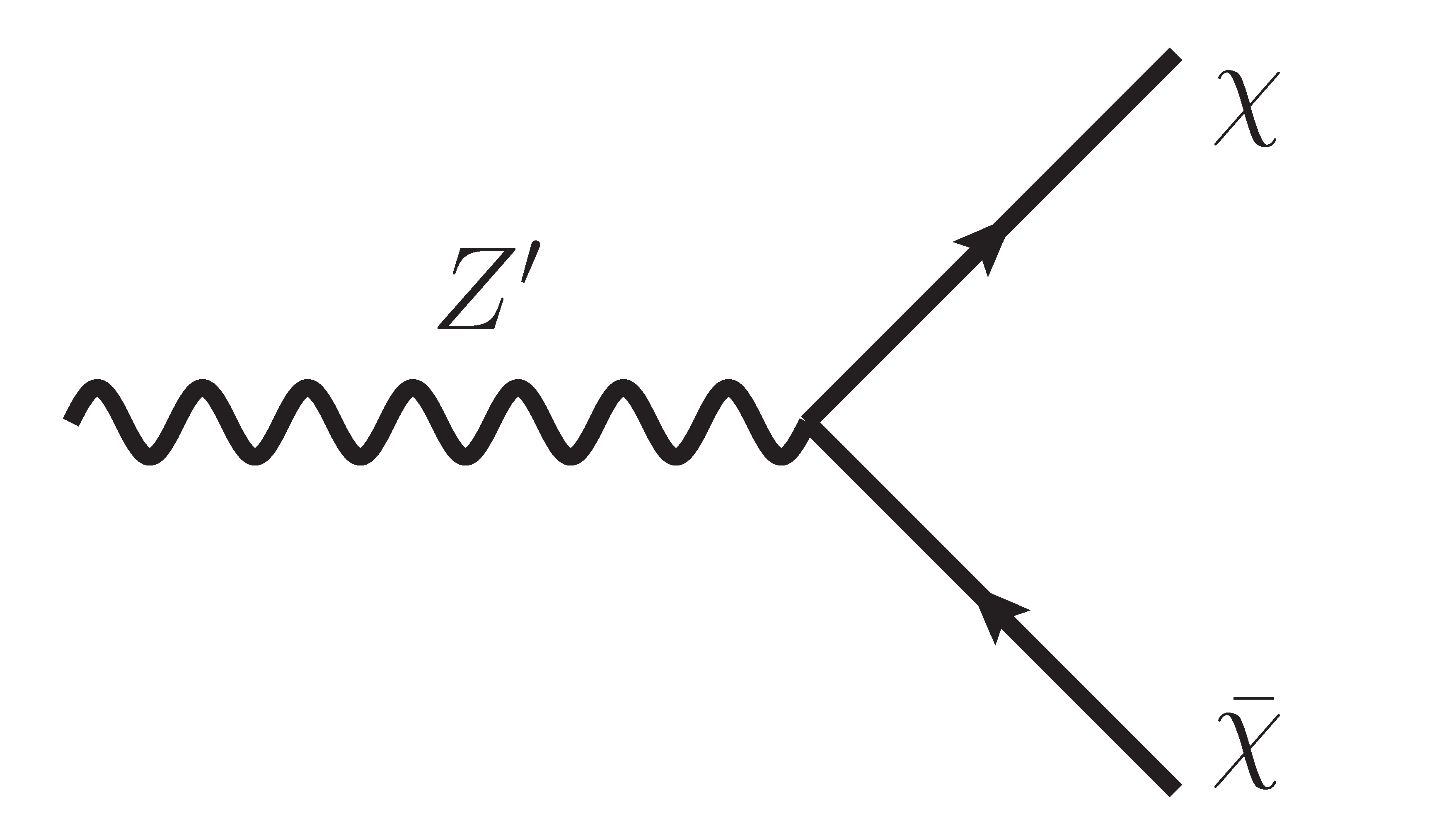}} \hspace{2cm}
 \subfloat[]{\includegraphics[width=0.4\textwidth]{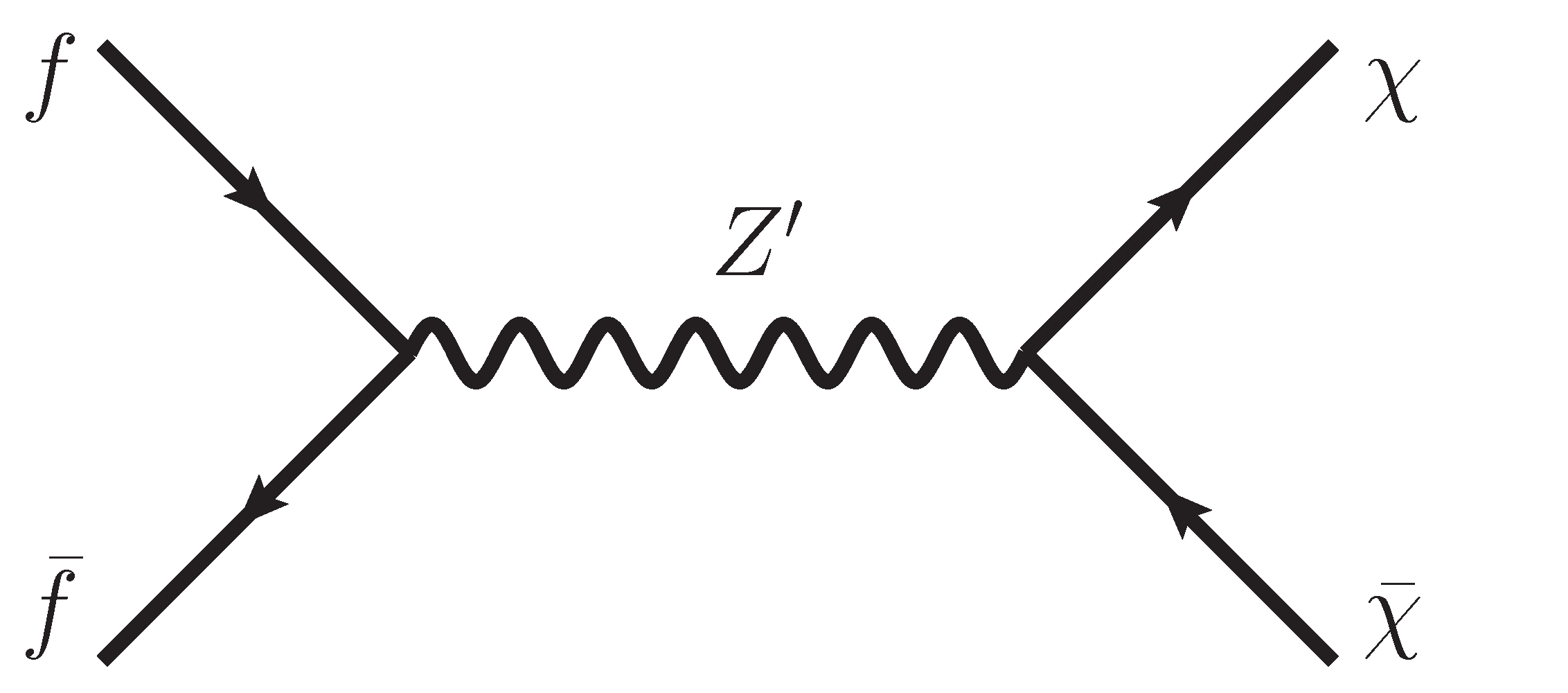}}
  \caption{Representative Feynman diagrams contributing to the dark fermion abundance in our $Z'$ model, (a) via decay of the mediator $Z'$ (b) via $2 \to 2$ production from the bath. 
   \label{fig:two-mediator-F-Graphs}}
\end{center}
\end{figure}

As a first illustrative example, we consider the case of a $Z'$ portal (an even particle) with pure-vector couplings to the SM and the (Dirac fermion) DM, the latter being the only odd particle that belongs to ${\cal F}$. This is much like in the so-called Simplified Models of dark matter often considered in the literature \cite{Abercrombie:2015wmb}. The Lagrangian is given by:
\beq
{\cal L}_{\rm int} = - g_\chi Z'_\mu \bar\chi\gamma^\mu \chi - \sum_f g_f Z'_\mu \bar f\gamma^\mu f
\label{eq:Z'}
\eeq
where the sum runs over the SM fermions. The model is anomaly-free if the various $g_f$ couplings are related through $g_e = g_\nu = -3 g_u = -3 g_d \equiv -3 g_q$, and we take these couplings to be universal between generations. We further assume that the Higgs doublet is neutral under $U(1)'$. Given these conditions, the model is characterised by two couplings, $g_\chi$ and $g_q$, and two masses, $m_{Z'}$ and $m_\chi$. Moreover, SM bosons do not participate in the production of DM, thus, as we will see below, the effect of statistics to the final DM yield is not very important. The diagrams contributing to dark matter production via $Z'$ decay or $2\to 2$ annihilation are shown in Fig.~\ref{fig:two-mediator-F-Graphs}.

The different regimes that could arise in freeze-in production through a portal particle \cite{Chu:2011be} can be classified according to the respective values of the two masses, as well as the assumed value of the reheating temperature $T_R$, as\footnote{In this discussion we do not consider situations in which the DM is heavier than the reheating temperature such as, \textit{e.g.}, in the case of WIMPzillas \cite{Chung:1998zb}.}:
\begin{itemize}
\item {\it Off-shell regime}, with $m_{Z'} < 2m_\chi < T_R$. In this case the DM production always occurs through $2\to 2$ annihilations $f\bar f\to\chi\bar\chi$ via an off-shell mediator.
\item {\it On-shell regime}, with $2m_\chi < m_{Z'} < T_R$. The DM production will be dominated by the resonant production $f\bar f\to Z'\to \chi\bar\chi$, \textit{i.e.} it is effectively a decay process. 
\item {\it EFT regime}, with $2m_\chi < T_R < m_{Z'}$. In this regime, and since the DM production is assumed to start at $T=T_R$ (for a few exceptions to this rule \textit{cf e.g.} \cite{Garcia:2017tuj,Chen:2017kvz}), the $Z'$ can be integrated out and the $2\to 2$ process effectively shrinks to a point-like interaction. 
\end{itemize}  

Simple dimensional arguments together with the behaviour of the $2\to 2$ cross section can provide intuition about the period during which the DM production mostly happens. In terms of the DM production yield $Y_\chi$, Eqs.~\eqref{yieldDM} and~\eqref{coll2to2} imply
\beq
Y_\chi \sim M_{\rm Pl} \int \frac{dT}{T^5} \int ds~s^{3/2}~\sigma(s)~\tilde K_1(\sqrt{s}/T,x_1,x_2,0,\eta_1,\eta_2)~,
\label{eq:Yhandy}
\eeq
where we have omitted the $g^*,g^*_s$ terms as well as any numerical prefactor. Substituting in Eq.~\eqref{eq:Yhandy} the $s$-dependence of the cross section that corresponds to each regime leads to the expressions shown in Table \ref{tab:regimes}. 

\begin{figure}
  \centering
  \includegraphics[scale=0.5]{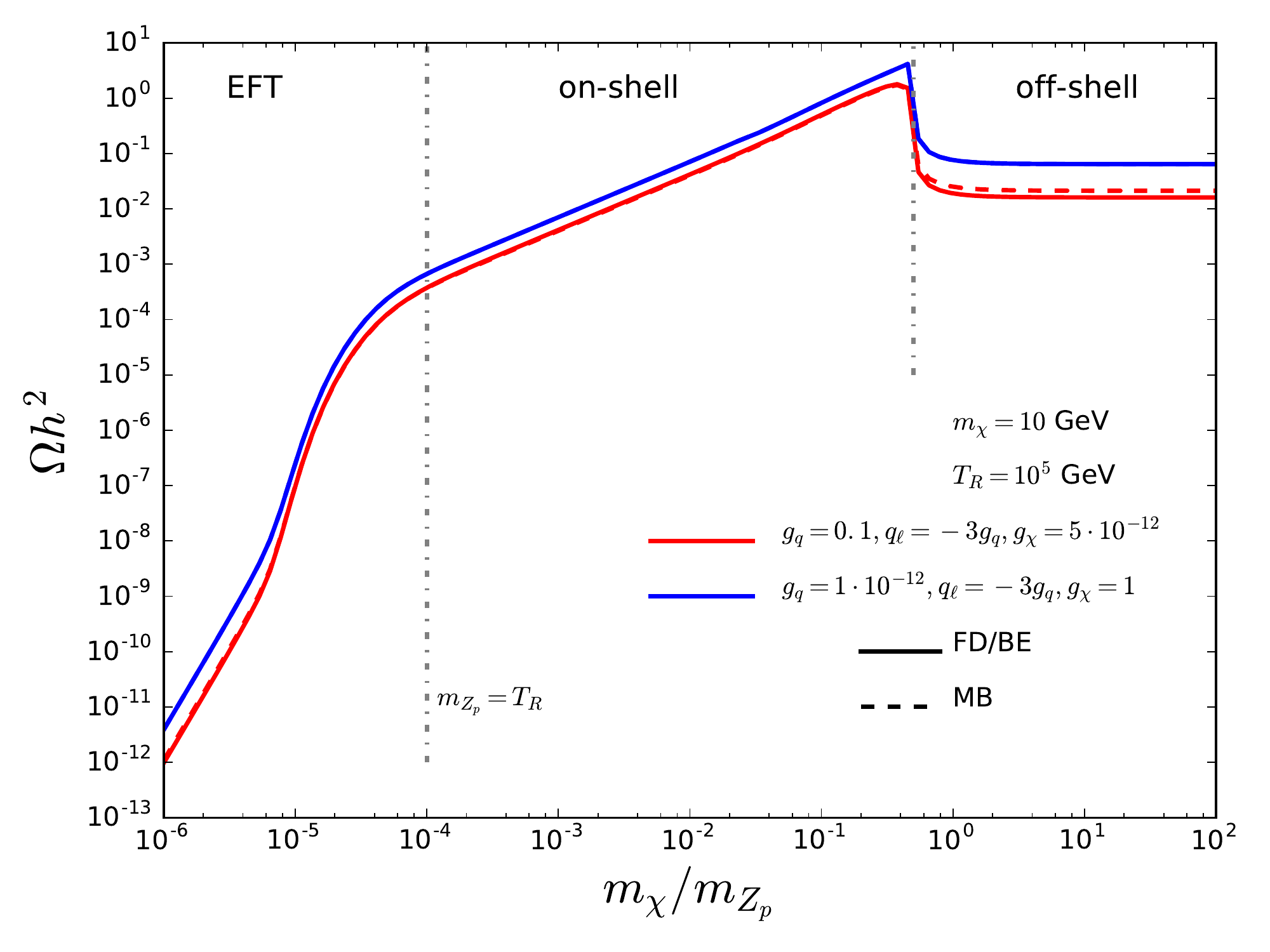} 
  \caption{Relic abundance as a function of the mass ratio $m_\chi/m_{Z'}$, for two coupling choices $(g_\chi,g_q) = (5\cdot 10^{-12}, 0.1)$ -- red -- and $(1, 10^{-12})$ -- blue -- with solid lines taking into account the exact distribution functions of the SM fermions. Dashed lines instead assume the Maxwell-Boltzmann approximation (see text for details). The dark matter mass and reheating temperature have been fixed to 10 GeV and $10^5$ GeV, respectively. The gray dot-dashed vertical lines at $m_\chi/m_{Z'}=10^{-4}$ and $m_\chi/m_{Z'}=0.5$ mark respectively the boundaries of the 3 different production regimes.}
\label{fig:Zp}
\end{figure}

The full numerical results are shown in Fig.~\ref{fig:Zp} for a fixed DM mass, $m_\chi=10$ GeV, and reproduce the expected behaviour. The red (blue) lines correspond to the case in which the mediator couples more strongly to the visible (dark) sector, whereas the solid (dashed) lines assume Fermi-Dirac (Maxwell-Boltzmann) statistics for the SM fermions.

\begin{table}
\begin{center}
\begin{tabular}{|c||c|c|}
\hline
regime & $\sigma(s)$ & $Y_\chi$ \\
\hline
off-shell & $\frac{g_f^2 g_\chi^2}{s}$ & $g_f^2 g_\chi^2 /m_\chi$\\
\hline 
on-shell & $\frac{g_f^2 g_\chi^2 m_{Z'}}{\Gamma}\delta(s-m^2_{Z'})$ & $ g_f^2 g_\chi^2 /\Gamma $\\
\hline
EFT & $g_f^2 g_\chi^2 s/m^4_{Z'}$ & $g_f^2 g_\chi^2 T^3_R/m^4_{Z'}$\\
\hline
\end{tabular}
\end{center}
\caption{Dependence of the $f\bar f \to\chi\bar\chi$ cross section (middle column) and yield (right column) on the parameters of the model, where $\Gamma$ is the total $Z'$ decay width. All numerical factors and effective degrees of freedom are not shown. For the on-shell regime, the Narrow Width Approximation is assumed for the sake of illustration.
}
\label{tab:regimes}
\end{table}

Some comments are in order. The general features of Fig.~\ref{fig:Zp} are straigthforward to understand, recalling that the relic abundance of DM scales as $\Omega_\chi \propto m_\chi Y_\chi$: deep inside the EFT regime we obtain the $1/m_{Z'}^4$ behaviour expected from Table \ref{tab:regimes}, whereas the off-shell regime (the right-most part of the figure) features a plateau, since the relic abundance effectively does not depend on the mediator mass\footnote{In this regime there is no dependence on the DM mass either (modulo the dependence coming from the effective degrees of freedom), because the $m_\chi$ factor appearing in the expression for $\Omega_\chi$ is compensated by the one appearing in the yield (see Table \ref{tab:regimes}).}. Finally, in the on-shell regime, decreasing the mediator mass (\textit{i.e.} moving towards the right-hand side of the figure) leads to a  decrease of  the decay width, and thus  an increase in  $\Omega_\chi$ until the on-shell production is no longer possible because of kinematical reasons. We have checked that for other values of the DM mass and the reheating temperature these features remain unaltered, as expected.  
An exact treatment of the statistical distributions of bath particles (difference between solid and dashed lines for each coupling choice) generally  leads to a decrease in the relic density of about 4\% in the EFT regime and of 25\% in other regimes. One exception is found in the on-shell regime for the case where the $Z'$ couples mostly to the SM fermions. In this case, the use of Fermi-Dirac distributions rather leads to a mild increase (4\%) in the relic density, which is due to the fact that the mediator production suppression characterising Fermi statistics is compensated by a smaller decay width $\Gamma$ (see Table~\ref{tab:regimes}). We discuss next other models where bosons participate in the DM production, such that the effect of statistics becomes more important.
 
\subsection{Scalar portal}\label{sec:scalarportal}

\begin{figure}
  \centering
  \includegraphics[scale=0.5]{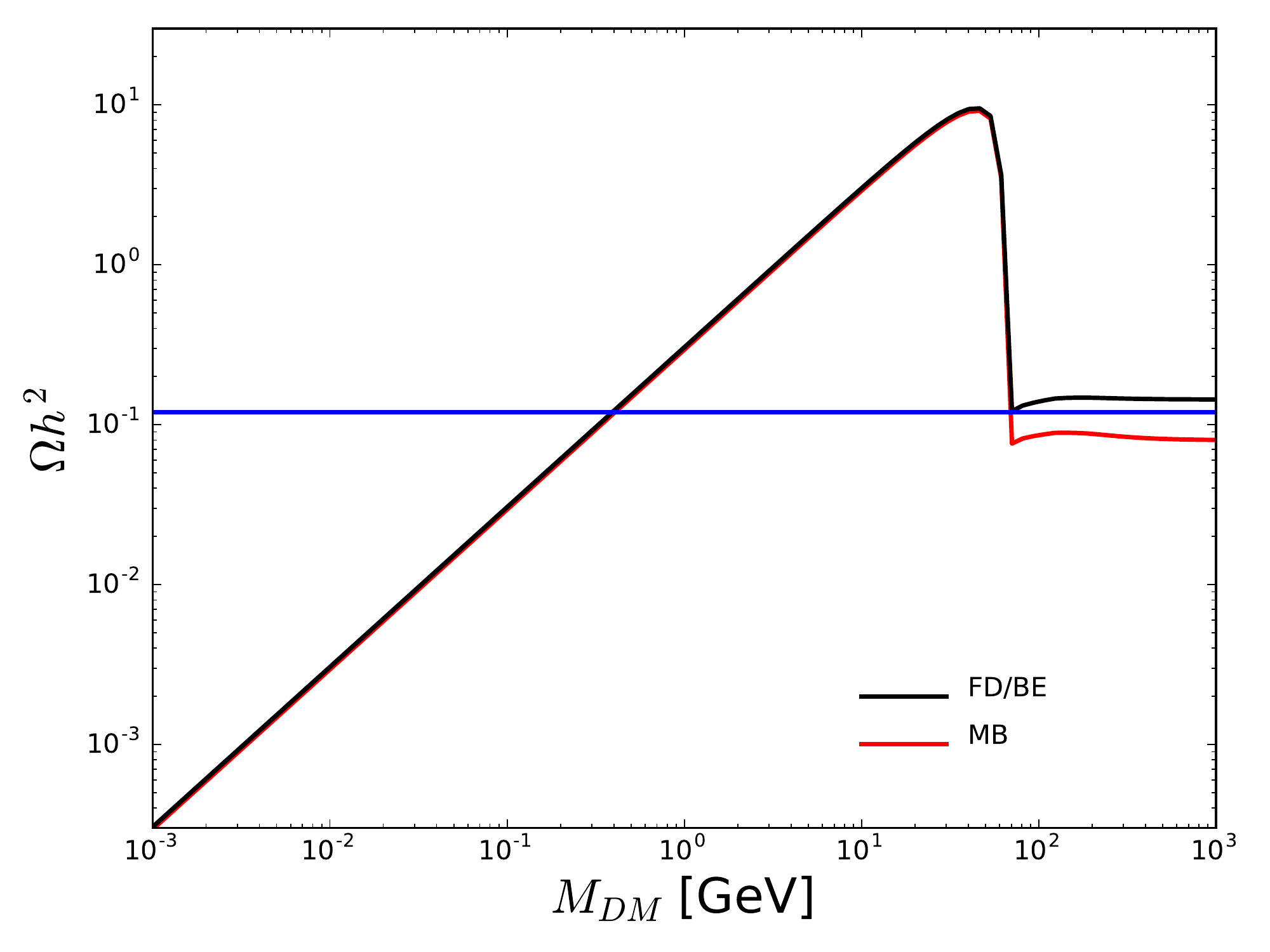} 
  \caption{Comparison of the relic density in the singlet scalar DM model assuming Maxwell-Boltzmann (red) and Bose-Einstein/Fermi-Dirac statistics (black), for $\lambda_{hs}=1 \times10^{-11}$. The blue line depicts the DM abundance determined by PLANCK. In this figure we take $T_R=10^6$ GeV.}
\label{fig:singlet}
\end{figure}

The singlet scalar model is  a minimal extension of the SM containing only an additional real singlet scalar field which is odd under a ${\cal{Z}}_2$ symmetry~\cite{McDonald:2001vt}. This field only possesses direct couplings to the SM Higgs boson, as described by the Lagrangian 
\begin{equation}\label{eq:singletdoubletmediator}
{\cal{L}}  = {\cal{L}}_{\rm SM} 
 + \frac{1}{2} \left(\partial_\mu S\right)\left(\partial^\mu S\right) + \frac{\mu_s^2}{2} S^2 - \frac{\lambda_s}{4} S^4 - \lambda_{hs} \left(H^\dagger H\right) S^2 
\end{equation}
and can be a viable dark matter candidate. The two free parameters of the model can be chosen as the singlet mass, $m_S^2=\mu_s^2 + \lambda_{hs} v^2$ and the coupling $\lambda_{hs}$. In this model the mediator is the SM Higgs boson. When $S$ is light enough, $m_S<m_h/2$, then in the freeze-in regime (that is for $\lambda_{hs} << 1$), DM is mainly produced from the decay of the Higgs. In this case  the relic density scales  proportionally to the DM mass, $\Omega h^2 \propto \lambda_{hs}^2 m_S$, as shown in Fig.~\ref{fig:singlet}. Here,  effects from statistics are mild (of the order of 3.5\% as mentioned in section~\ref{sec:decay_thermal}).
When the mediator cannot be produced and decay on-shell, the DM abundance is substantially suppressed. All SM particles contribute to DM production, including $W$'s, $Z$'s and $h$'s. Thus, we expect a significant difference between the results obtained assuming a Maxwell-Boltzmann or the full Bose -Einstein distribution for the bath particles. Indeed, we find nearly a factor 2 increase in the DM density, as shown in Fig.~\ref{fig:singlet}. Note that for the same value of the coupling, the relic density constraint can be satisfied in two regimes: for light DM (sub-GeV) and for weak scale DM. These results are in good agreement with the ones obtained in previous studies of this model~\cite{Yaguna:2011qn}, if a Maxwell-Boltzmann distribution is assumed.

\subsection{Extended dark sector}\label{sec:t-channel}

As a final example, we study a simple model in which the ``dark sector'' consists of more than one particle. This model could be of interest for collider searches for long-lived particles. We consider an extension of the Standard Model by an additional real scalar field $s$ that transforms trivially under $SU(3)_c \times SU(2)_L \times U(1)_Y$ as well as an additional vector-like charged lepton $E$ transforming as $\left( \mathbf{1}, \mathbf{1}, \mathbf{-1} \right)$
\footnote{The vector-like nature of $E$ ensures that the model is anomaly-free.}. Both particles are taken to be odd under a discrete ${\cal{Z}}_2$ symmetry, whereas all Standard Model fields are taken to be even. Under these assumptions, the Lagrangian of the model reads
\begin{align}\label{eq:extDSlag}
{\cal{L}} & = {\cal{L}}_{\rm SM} + \left(\partial_\mu s\right)\left(\partial^\mu s\right) + \frac{\mu_s^2}{2} s^2 - \frac{\lambda_s}{4} s^4 - \lambda_{hs} s^2 \left(H^\dagger H\right) \\ \nonumber
& + i \left( \bar{E}_L \slashed{D} E_L  + \bar{E}_R \slashed{D} E_R \right) - \left( m_{E} \bar{E}_L E_R + y_{s} s \bar{E}_L e_R + {\rm{h.c.}} \right) 
\end{align}
where $E_{L,R}$ and $e_R$ are the left- and right-handed components of the heavy lepton and the right-handed component of the Standard Model electron respectively and for simplicity we have neglected couplings to the second and third generation leptons. The model is described by five free parameters, namely
\begin{equation}
\mu_s, \ \lambda_s, \ \lambda_{hs}, \ m_E, \ y_{s}
\end{equation}
out of which $\lambda_s$ is irrelevant for our purposes\footnote{In some cases dark matter self-couplings can be relevant for the computation of the DM abundance, \textit{cf} \cite{Bernal:2015ova,Bernal:2015xba}. We do not consider these possibilities here.} whereas $\mu_s$ can be traded for the physical mass of $s$ through
\begin{equation}
\mu_s^2 = -m_s^2 + \lambda_{hs} v^2
\end{equation}
where $v$ is the Higgs field vacuum expectation value. For simplicity, we will also take the coupling $\lambda_{hs}$ to be identically zero, as Higgs portal-like interactions were already studied in section \ref{sec:scalarportal}. These choices leave us with only three free parameters
\begin{equation}
m_s, \ m_E, \ y_{s} \ .
\end{equation}
For $m_s < m_E$, the scalar $s$ becomes stable and can play the role of a dark matter candidate. 
\begin{figure}
 \begin{center}
 \subfloat[]{\includegraphics[width=0.2\textwidth]{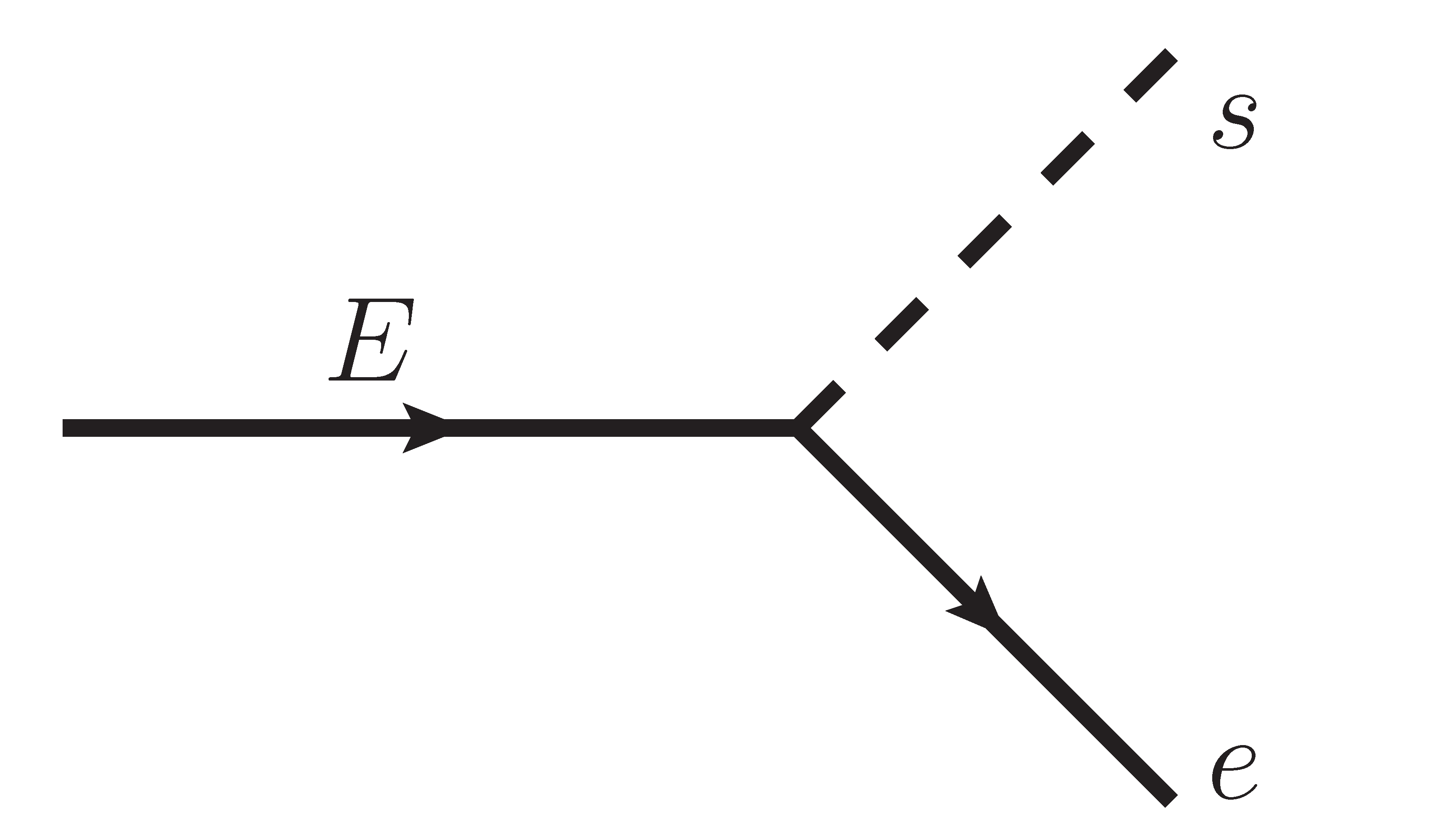}} \hspace{1cm}
 \subfloat[]{\includegraphics[width=0.2\textwidth]{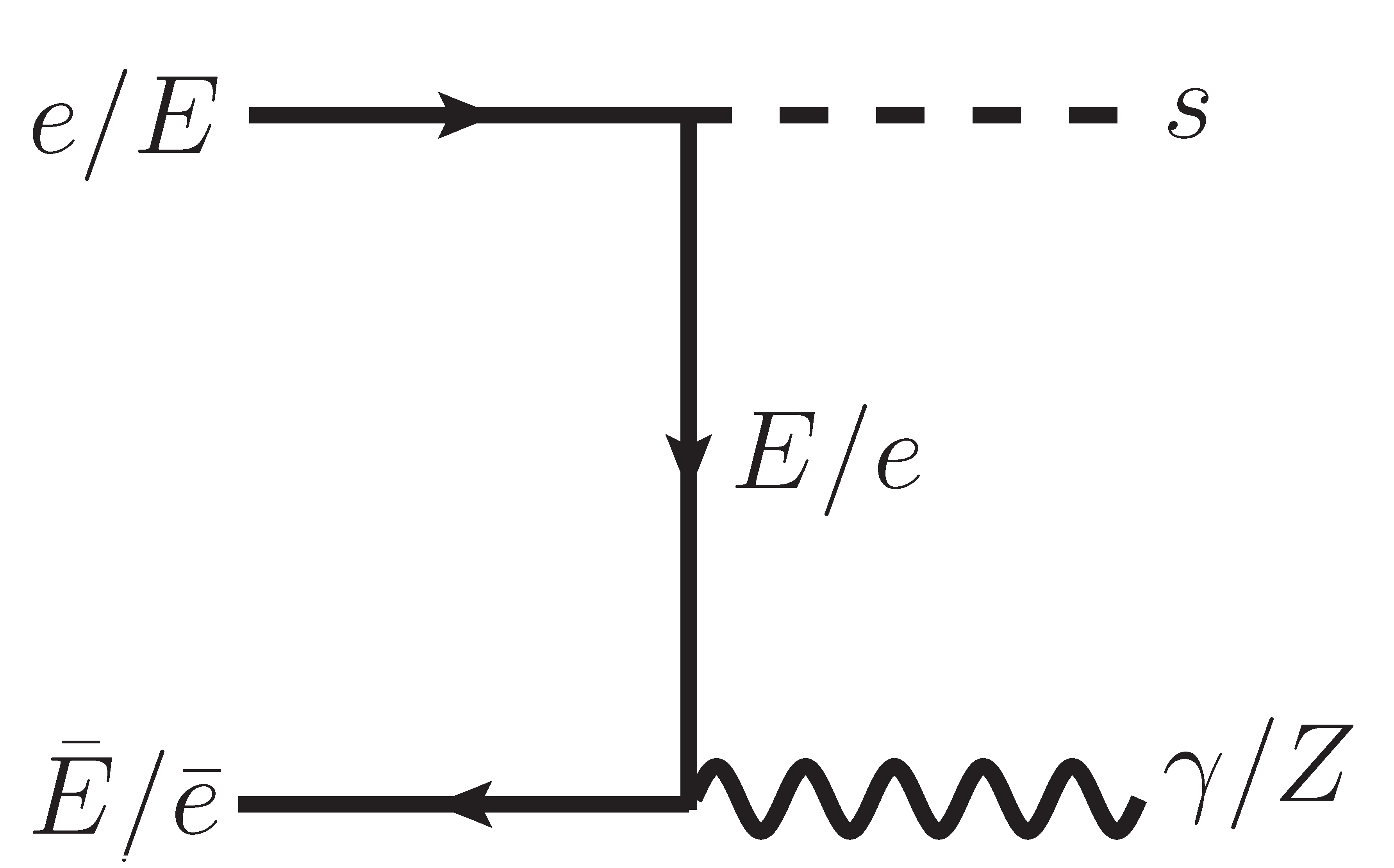}}\hspace{1cm}
 \subfloat[]{\includegraphics[width=0.2\textwidth]{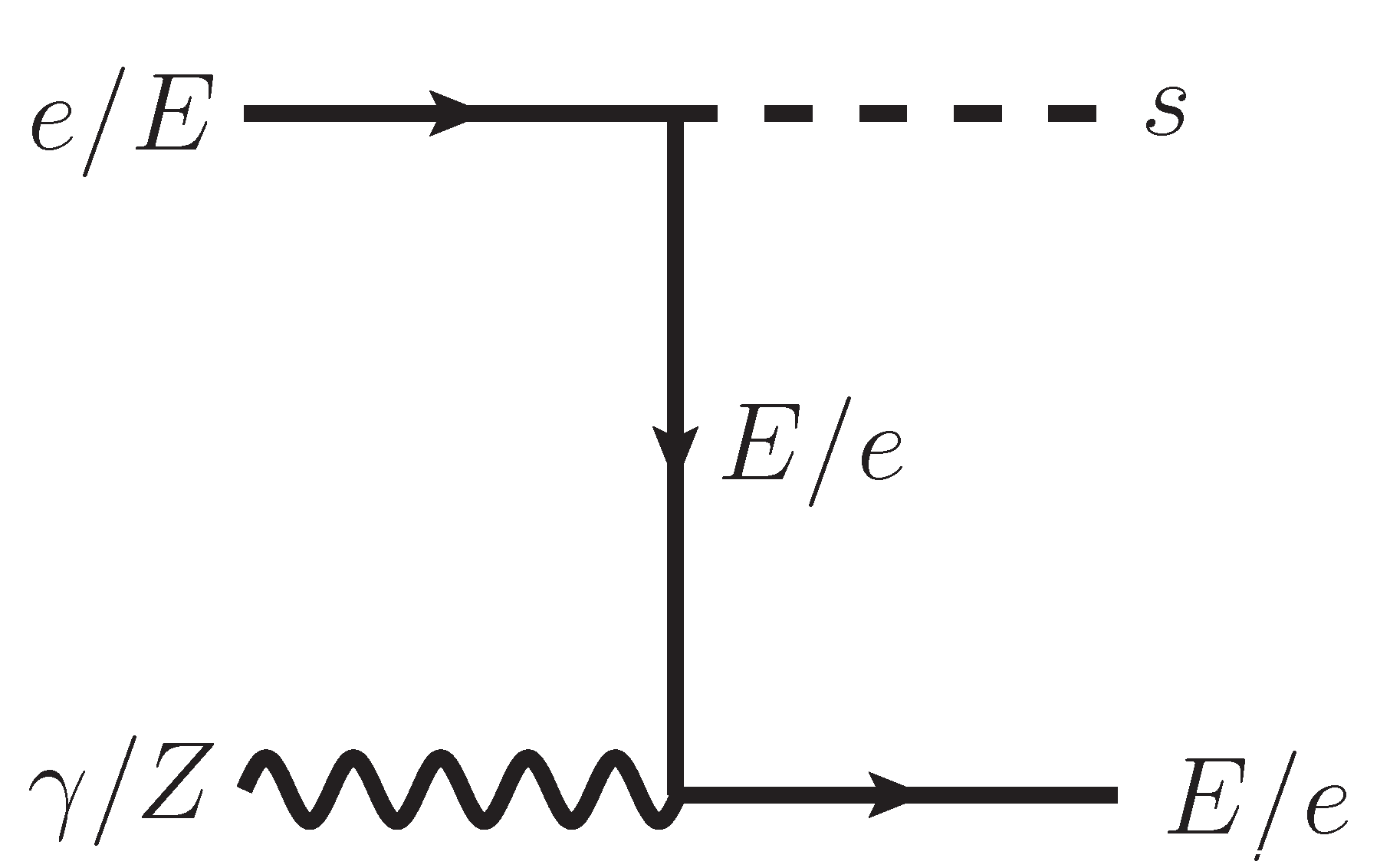}}\\
 \subfloat[]{\includegraphics[width=0.2\textwidth]{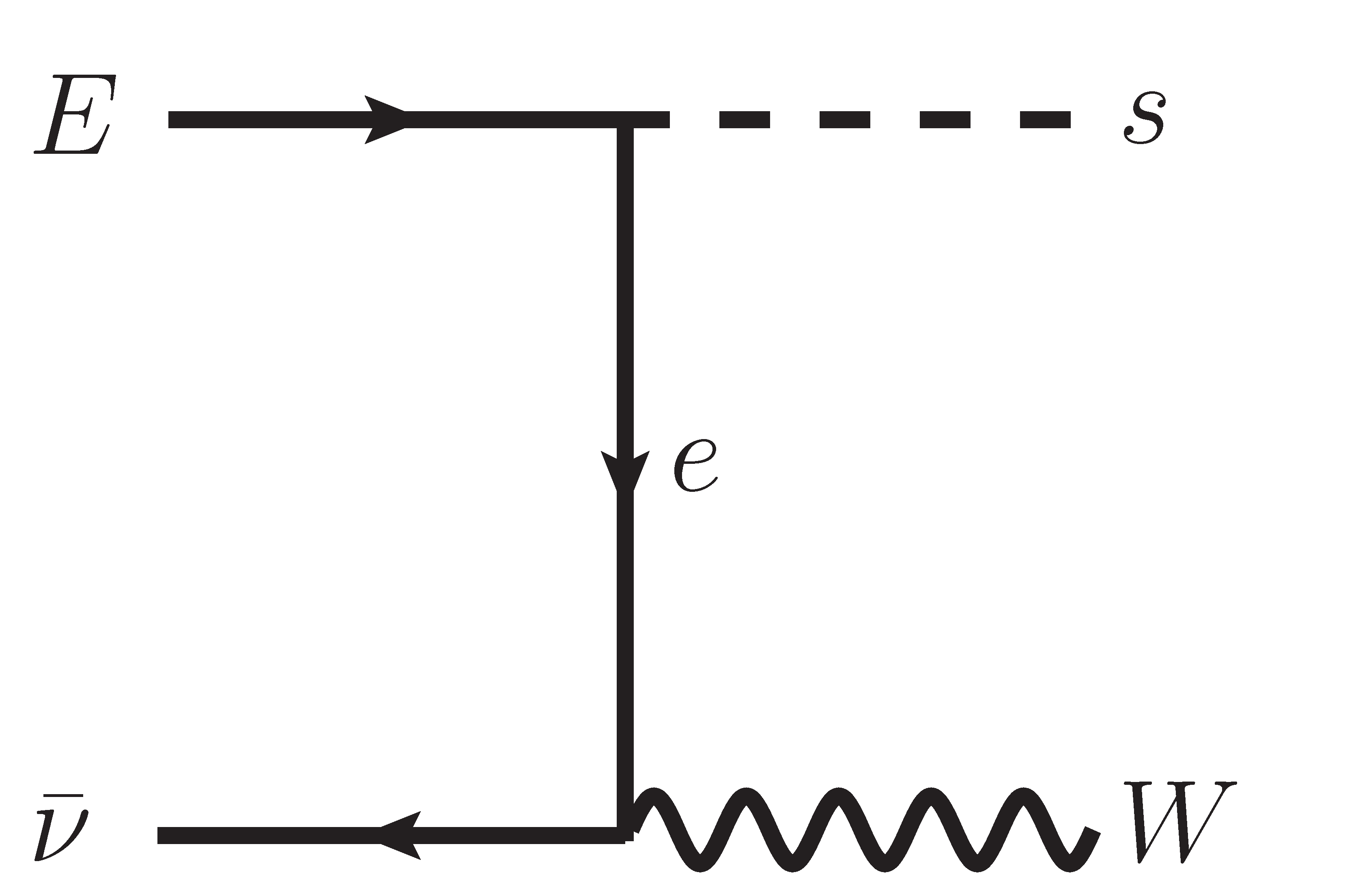}} \hspace{2cm}
 \subfloat[]{\includegraphics[width=0.2\textwidth]{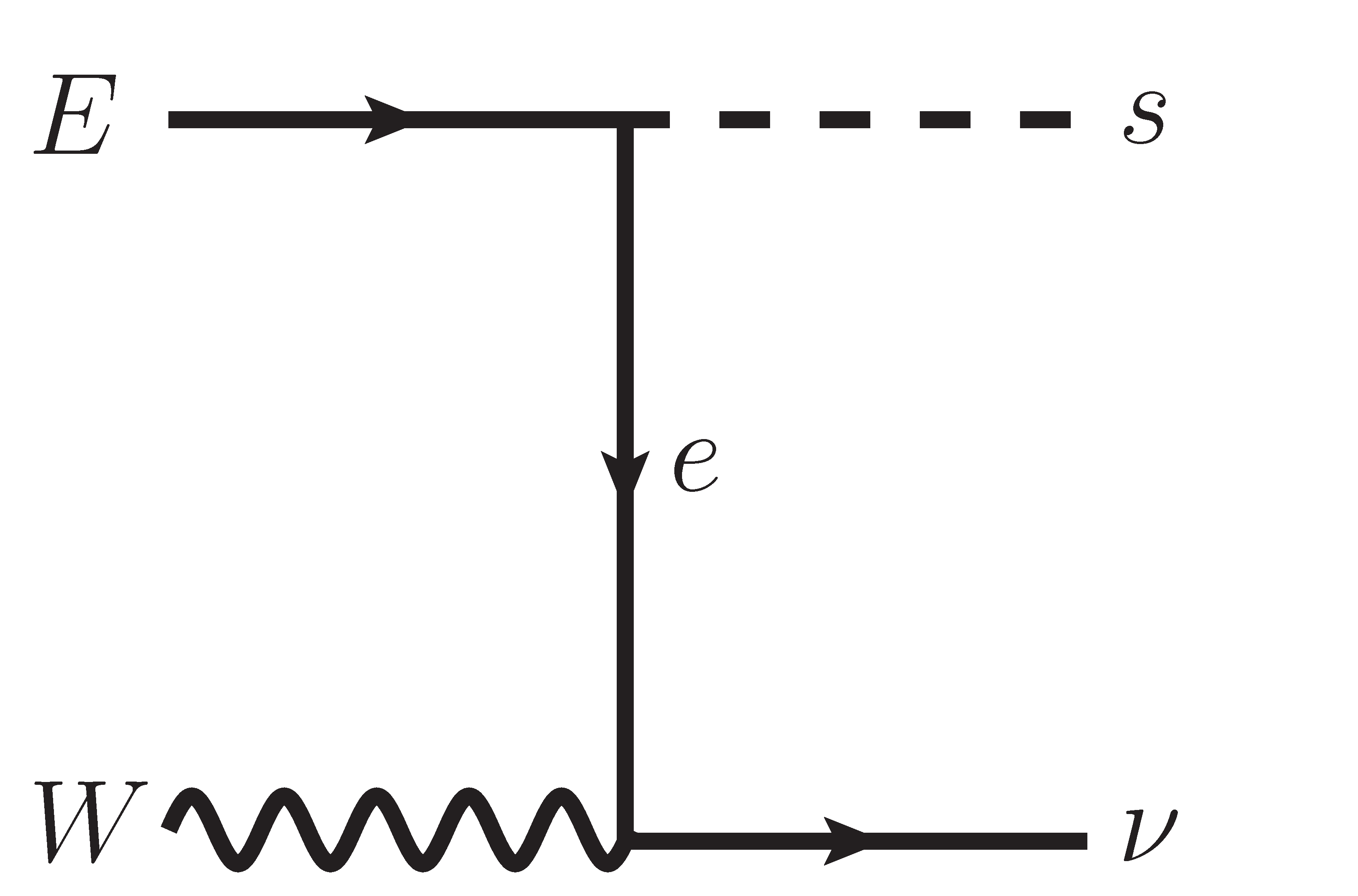}}\\
  \caption{Representative Feynman diagrams contributing to the dark fermion abundance in our extended dark sector model, (a) via decay of the Heavy lepton $E$ or (b)--(e) via $2 \to 2$ production from the bath. 
   \label{fig:F-Graphs-extendedDS}}
\end{center}
\end{figure}
\\
\\
In this scenario, and given our choice of neglecting the Higgs portal interaction, there are two types of processes that contribute to the dark matter abundance, depicted in Fig.~\ref{fig:F-Graphs-extendedDS}. First, $s$ can be produced through the decay of the heavy lepton, $E \rightarrow s + e$. Secondly, through annihilation processes involving the $t$-channel exchange of an ordinary or heavy electron. Note that the heavy lepton is kept in thermal equilibrium with the Standard Model plasma due to its gauge interactions. In order to carry out our numerical analysis, we have implemented the model described by the Lagrangian in Eq.~\eqref{eq:extDSlag} in the {\tt FeynRules} package \cite{Alloul:2013bka} in order to generate the necessary {\tt CalcHEP} \cite{Belyaev:2012qa} model files that can be subsequently used by \micro.

\begin{figure}[ht]
  \centering
  \includegraphics[scale=0.5]{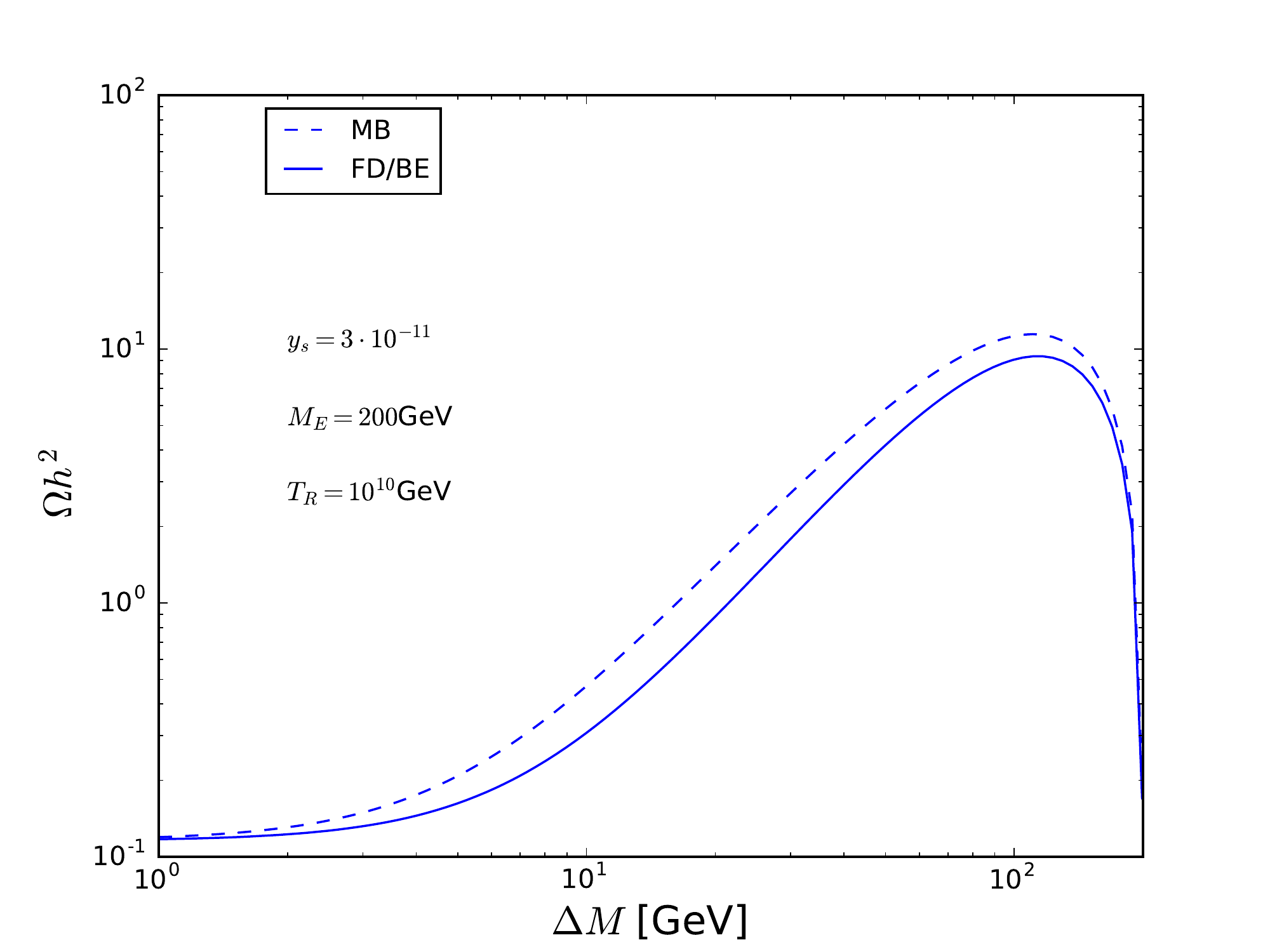} 
  \caption{Relic density of the dark scalar in our extended dark sector model, assuming Maxwell-Boltzmann (solid) and Bose-Einstein/Fermi-Dirac statistics (dashed) for $y_s=3 \times10^{-11}$.  Here $\Delta M$ is the mass difference between the heavy lepton and the scalar FIMP and $T_R=10^{10}$ GeV. }
\label{fig:tchannel}
\end{figure}

Our results are shown in Fig.~\ref{fig:tchannel}, where we show the predicted DM abundance as a function of the dark scalar - heavy lepton mass splitting, for one particular choice of the heavy lepton mass and its coupling to DM, assuming $T_R = 10^{10}$ GeV. In the same figure, we moreover illustrate the effect of an exact treatment of the statistical distributions of bath particles (dashed vs solid line for Maxwell-Boltzmann vs Fermi-Dirac/Bose-Einstein statistics respectively). Our numerical analysis reveals that for large values of the mass splitting (\textit{i.e.} towards the right hand-side of the figure), the dominant contribution to the DM abundance arises from decays of the heavy leptons. As the mass splitting becomes smaller, however, scattering processes take over. The effect of statistics is more pronounced in the regime where decay processes dominate and for moderate values of the mass splitting. As we observe, the use of Fermi-Dirac distributions can lead to a significant decrease in the relic density. Indeed, in this case the energy of the  final SM fermion $E \ll T\approx m_Y/3$, such that  the factor that enters Eq.~\eqref{eq:coll12dec} $(1-f)\propto (e^{E/T}+|\eta|)^{-1} \rightarrow 1/2$ when $|\eta|\approx 1$. Incidentally, for the parameter choices adopted here we see that the observed DM abundance in the Universe can be obtained in the regime where scattering dominates, although a different choice of parameters (in particular, decreasing the coupling $y_s$) would change this picture.

\section{Installation and sample output}
\label{sec:install}

The code can be downloaded from~\cite{micro-hp}. After unpacking, the user must go to the \verb|micromegas_5.0| directory and type\\

\verb|make| ~or~ \verb|gmake|\\

\noindent
To work with one of the models already included in the distribution, go to the relevant model directory, for example  \\

\verb|cd SingletDM|\\

\noindent
and compile the main code with\\

\verb|make main=main.c| \\

\noindent
This sample code contains the following lines to compute the relic density via decay or scattering processes,

\begin{verbatim}
  double TR=1E6;
  toFeebleList("~x1");
  omegaFi=darkOmegaFi(TR,&err);
  printChannelsFi(0,0,stdout);
\end{verbatim}

\noindent
Running the code   with the sample data file\\
\verb|./main data1.par| \\

\noindent
will  lead to the following output

\begin{verbatim}
=== MASSES OF HIGGS AND ODD PARTICLES: ===
Higgs masses and widths
      h   125.00 3.07E-03

Masses of odd sector Particles:
~x1  : Mdm1  =     1.0 || 

==== Calculation of relic density =====
Freeze-in Omega h^2 =1.220E+00
# Channels which contribute to omega h^2 via freeze-in
 6.680E-01  B, b ->  ~x1,~x1 
 2.159E-01  G, G ->  ~x1,~x1 
 7.289E-02  L, l ->  ~x1,~x1 
 2.745E-02  C, c ->  ~x1,~x1 
 6.335E-03  A, A ->  ~x1,~x1 
 3.570E-03  W-, W+ ->  ~x1,~x1 
 2.773E-03  S, s ->  ~x1,~x1 
 1.362E-03  Z, Z ->  ~x1,~x1 
 1.336E-03  h, h ->  ~x1,~x1 
 2.582E-04  M, m ->  ~x1,~x1 
 1.221E-04  T, t ->  ~x1,~x1 
 6.933E-06  U, u ->  ~x1,~x1 
 6.933E-06  D, d ->  ~x1,~x1 

h :   total width=3.067960E-03 
 and Branchings:
3.085202E-03 h -> A,A
1.051279E-01 h -> G,G
2.985241E-04 h -> m,M
8.427125E-02 h -> l,L
8.015904E-06 h -> u,U
8.015904E-06 h -> d,D
3.173942E-02 h -> c,C
3.206313E-03 h -> s,S
7.722552E-01 h -> b,B
2.517642E-18 h -> ~x1,~x1
\end{verbatim}

\section{Discussion and conclusions}
\label{sec:conclusion}

Freeze-in constitutes a viable mechanism for the generation of the observed dark matter abundance in the Universe. Although the corresponding Boltzmann equation for dark matter appears to be simpler than its freeze-out counterpart, in practice the loss of equilibrium at high temperatures can introduce several complications when working within general extensions of the Standard Model of particle physics. 

First, in typical freeze-out models, the equilibrium condition erases all memory concerning processes that occur at high temperatures. In freeze-in, conversely, DM production can even be \textit{dominated} by these temperatures, which implies that its density evolution may have to be tracked back in time up to $T_R$\footnote{Arguably, such scenarios are at odds with the original spirit of freeze-in as an IR-dominated mechanism~\cite{McDonald:2001vt,Hall:2009bx}. Indeed, the DM abundance in UV-dominated freeze-in scenarios depends on an additional parameter, the reheating temperature $T_R$. Nevertheless, from a thermodynamical standpoint, the two cases do not present substantial differences. In this spirit, throughout this work we have collectively used the term ``freeze-in'' for both cases.}. On a related topic, with the exception of coannihilation and excluding possibilities such as late entropy injection from decays of long-lived particles, standard freeze-out scenarios are relatively insensitive to the behaviour of particles heavier than the  DM itself: equilibrium is restored extremely fast in the early Universe and, down to temperatures around $m_\chi/T \sim 20$, dark matter simply traces its equilibrium distribution. This is certainly \textit{not} the case in the freeze-in scenario, in which one has to keep track of the contribution  to DM production of all particles in the spectrum. In particular, it is fairly easy to envisage models in which DM is not the only FIMP in the spectrum, which in turn implies that an additional Boltzmann equation has to be written down for each feebly coupled degree of freedom and solved simultaneously with the DM one. Moreover, while in freeze-out the initial DM abundance is fairly irrelevant for the computation of its present-day abundance, in freeze-in the initial condition has to be supplied by hand, at least from a low-energy standpoint. Last but not least, although in freeze-out the fermionic or bosonic nature of bath particles can be neglected when it comes to their phase-space distributions, in freeze-in it becomes relevant and may substantially alter the predicted dark matter abundance.

In this paper we presented a new version of the \micro~dark matter code to compute the abundance of general feebly coupled dark matter candidates according to the freeze-in mechanism. We presented the formalism that is necessary in order to treat different ``types'' of freeze-in depending on the nature of the particles involved in the dark matter production process, and to tackle the complications described above. We recovered several well-known results in the literature, such as the fact that in infrared-dominated freeze-in scenarios dark matter production through decays of a mediator $Y$ peaks around $m_Y/3$, and quantified the extent to which adopting a Maxwell-Boltzmann distribution for the bath particles constitutes a good approximation. The latter was found to not always be the case and can lead to a misprediction of the dark matter abundance by up to a factor 2. Interestingly, we have moreover found that in the case of production through decays of a heavier state which is \textit{not} in chemical equilibrium with the SM thermal bath, \textit{i.e.} a particle that freezes-in itself, the predicted DM abundance does not depend crucially on whether or not the heavier particle is in kinetic equilibrium. However, since we cannot make general statements on whether or not this result holds in full generality, in \micro~we provide two computational methods which can be used to estimate its validity. Finally, we have carefully taken into account the possibility that DM might be produced at a temperature $T_d$ through the decay of a resonance which was itself produced at some temperature $T_c \gg T_d$, by properly modifying the mediator width entering the $2\to 2$ scattering cross-section. 

One limitation of the code is that it does not check whether the DM couplings comply with the freeze-in assumption, however the same can be said of  all DM public codes  that compute the DM abundance according to the freeze-out mechanism: for sufficiently high temperatures, thermal equilibrium between the dark and visible sectors is always assumed regardless of the input coupling values.  The intermediate regime between freeze-in and freeze-out, which is arguably the most complicated one, has rarely been considered in the literature (for notable exceptions \textit{cf e.g.} \cite{Chu:2011be,Bae:2017dpt}) and involves the resolution of a more general form of the Boltzmann equation(s). It will be the topic of a future upgrade of \micro.

Despite the code limitations, we hope it will be a useful tool for theorists and experimentalists alike working on feebly coupled dark matter candidates, especially in light of the recent flourishing activity on long-lived particle searches at the LHC but also on searches for feebly coupled particles in intensity frontier experiments. We further hope that it will facilitate the study of freeze-in scenarios beyond the simplest options that have, for the most, so far been considered in the literature. 

\section{Acknowledgements}

We thank N. Bernal, M. Boudaud, M. Cirelli, M. Dutra, M. Heikinheimo, K. Jedamzik, A. Kusenko, Y. Mambrini, M. Pierre and K. Petraki for useful discussions. This work was supported in part by the French ANR, Project DMAstro-LHC ANR-12-BS05-0006,  the {\it Investissements d'avenir}, Labex ENIGMASS, and by the Research Executive Agency (REA) of the European Union under the Grant Agreement PITN-GA2012-316704 (``HiggsTools"). We acknowledge the Galileo Galilei Institute for Theoretical Physics for the hospitality during the completion of this work. A.G. was supported by the Labex ILP (reference ANR-10-LABX-63) part of the Idex SUPER, and received financial state aid managed by the Agence Nationale de la Recherche, as part of the programme Investissements d'avenir under the reference ANR-11-IDEX-0004-02. A.P. acknowledges partial support from SOTON Diamond Jubilee Fellowship as well as from Royal Society International Exchanges grant IE150682.

\appendix

\section{Treatment of $t$-channel propagators}

For freeze-in DM production, one needs to calculate cross sections at energies which are much larger than the masses of the particles involved in the process. This can lead to numerical instabilities especially for $2\rightarrow 2$ processes containing a $t$-channel propagator. For such a process, the  differential cross-section has a pole contribution 
\begin{equation}
     \frac{d \sigma}{ d \cos{\theta_{13}}} \approx    \frac{A}{(1-\cos{\theta_{13}} +\delta_{13})^l}  
\end{equation} 
where $\theta_{13}$ is the angle between incoming and outgoing particles in the center-of-mass frame, while   for large values of the  Mandelstam variable $s$
\begin{equation} 
  \delta_{13}\approx \frac{2}{s}\big( m_t^2  +\frac{(m_1^2-m_3^2)(m_2^2-m_4^2)}{s}\big) \;\;,
\end{equation}
where  $m_1,m_2$ $(m_3,m_4)$ are the masses of incoming (outgoing) particles and $m_t$ is the propagator mass. Typically $l=1$ for fermions and $=2$ for bosons.

If $\delta_{13}>10^{-10}$  such a singularity can be successfully integrated using \verb|double precision| floating point numbers, however for smaller values of
 $\delta_{13}$ this is no longer sufficient. To circumvent this difficulty, {\tt micrOMEGAs} analyses the Feynman diagrams contributing to each process. When $t$-channel diagrams are found,  {\tt micrOMEGAs} first calculates the cross section in the region  $\cos{\theta_{13}}< 1-10^{-10}$. The code  then
calculates the differential cross section at  several  points  with small scattering  angles   to  find  the coefficients $A$ and $C$ in  
\begin{equation}
 \frac{d\sigma}{d cos{\theta_{13}}}= C+ \frac{A}{(1-\cos{\theta_{13}}+\delta_{13})^l}~.
\end{equation}    
This approximate formula is then integrated symbolically. This method leads to stable results although it may fail in two cases. First the value of 
$\delta_{13}$ used corresponds to the smallest mass running in the $t$-channel, it is however possible in models with several $t$-channel diagrams that the leading contribution comes from the propagator with a larger mass. Second, it can happen that  a pole  has a negligible contribution at $\cos{\theta_{13}}\approx 1-10^{-10}$ while giving a noticeable contribution to the total cross section. This occurs if $l=2$ and $\delta_{13}<10^{-30}$. Both these issues will be addressed in a future release of \micro.

\bibliography{micro}

\providecommand{\href}[2]{#2}\begingroup\raggedright\begin{thebibliography}{10}

\bibitem{Hinshaw:2012aka}
{\bfseries WMAP} Collaboration, G.~Hinshaw {\em et~al.}, ``{Nine-Year Wilkinson
  Microwave Anisotropy Probe (WMAP) Observations: Cosmological Parameter
  Results},'' \href{http://dx.doi.org/10.1088/0067-0049/208/2/19}{{\em
  Astrophys. J. Suppl.} {\bfseries 208} (2013) 19},
\href{http://arxiv.org/abs/1212.5226}{{\ttfamily arXiv:1212.5226
  [astro-ph.CO]}}.

\bibitem{Ade:2015xua}
{\bfseries Planck} Collaboration, P.~A.~R. Ade {\em et~al.}, ``{Planck 2015
  results. XIII. Cosmological parameters},''
  \href{http://dx.doi.org/10.1051/0004-6361/201525830}{{\em Astron. Astrophys.}
  {\bfseries 594} (2016) A13},
\href{http://arxiv.org/abs/1502.01589}{{\ttfamily arXiv:1502.01589
  [astro-ph.CO]}}.

\bibitem{Bertone:2004pz}
G.~Bertone, D.~Hooper, and J.~Silk, ``{Particle dark matter: Evidence,
  candidates and constraints},''
  \href{http://dx.doi.org/10.1016/j.physrep.2004.08.031}{{\em Phys. Rept.}
  {\bfseries 405} (2005) 279--390},
\href{http://arxiv.org/abs/hep-ph/0404175}{{\ttfamily arXiv:hep-ph/0404175
  [hep-ph]}}.

\bibitem{Bertone:2010zza}
J.~Silk {\em et~al.}, {\em {Particle Dark Matter: Observations, Models and
  Searches}}.
\newblock 2010.
\newblock
\url{http://www.cambridge.org/uk/catalogue/catalogue.asp?isbn=9780521763684}.
\newblock

\bibitem{Aprile:2017iyp}
{\bfseries XENON} Collaboration, E.~Aprile {\em et~al.}, ``{First Dark Matter
  Search Results from the XENON1T Experiment},''
\href{http://arxiv.org/abs/1705.06655}{{\ttfamily arXiv:1705.06655
  [astro-ph.CO]}}.

\bibitem{Tan:2016zwf}
{\bfseries PandaX-II} Collaboration, A.~Tan {\em et~al.}, ``{Dark Matter
  Results from First 98.7 Days of Data from the PandaX-II Experiment},''
  \href{http://dx.doi.org/10.1103/PhysRevLett.117.121303}{{\em Phys. Rev.
  Lett.} {\bfseries 117} no.~12, (2016) 121303},
\href{http://arxiv.org/abs/1607.07400}{{\ttfamily arXiv:1607.07400 [hep-ex]}}.

\bibitem{Akerib:2016vxi}
{\bfseries LUX} Collaboration, D.~S. Akerib {\em et~al.}, ``{Results from a
  search for dark matter in the complete LUX exposure},''
  \href{http://dx.doi.org/10.1103/PhysRevLett.118.021303}{{\em Phys. Rev.
  Lett.} {\bfseries 118} no.~2, (2017) 021303},
\href{http://arxiv.org/abs/1608.07648}{{\ttfamily arXiv:1608.07648
  [astro-ph.CO]}}.

\bibitem{Amole:2017dex}
{\bfseries PICO} Collaboration, C.~Amole {\em et~al.}, ``{Dark Matter Search
  Results from the PICO-60 C$_3$F$_8$ Bubble Chamber},''
  \href{http://dx.doi.org/10.1103/PhysRevLett.118.251301}{{\em Phys. Rev.
  Lett.} {\bfseries 118} no.~25, (2017) 251301},
\href{http://arxiv.org/abs/1702.07666}{{\ttfamily arXiv:1702.07666
  [astro-ph.CO]}}.

\bibitem{Ahnen:2016qkx}
{\bfseries Fermi-LAT, MAGIC} Collaboration, M.~L. Ahnen {\em et~al.}, ``{Limits
  to dark matter annihilation cross-section from a combined analysis of MAGIC
  and Fermi-LAT observations of dwarf satellite galaxies},''
  \href{http://dx.doi.org/10.1088/1475-7516/2016/02/039}{{\em JCAP} {\bfseries
  1602} no.~02, (2016) 039},
\href{http://arxiv.org/abs/1601.06590}{{\ttfamily arXiv:1601.06590
  [astro-ph.HE]}}.

\bibitem{Abdallah:2016ygi}
{\bfseries H.E.S.S.} Collaboration, H.~Abdallah {\em et~al.}, ``{Search for
  dark matter annihilations towards the inner Galactic halo from 10 years of
  observations with H.E.S.S},''
  \href{http://dx.doi.org/10.1103/PhysRevLett.117.111301}{{\em Phys. Rev.
  Lett.} {\bfseries 117} no.~11, (2016) 111301},
\href{http://arxiv.org/abs/1607.08142}{{\ttfamily arXiv:1607.08142
  [astro-ph.HE]}}.

\bibitem{Giesen:2015ufa}
G.~Giesen, M.~Boudaud, Y.~Genolini, V.~Poulin, M.~Cirelli, P.~Salati, and P.~D.
  Serpico, ``{AMS-02 antiprotons, at last! Secondary astrophysical component
  and immediate implications for Dark Matter},''
  \href{http://dx.doi.org/10.1088/1475-7516/2015/09/023,
  10.1088/1475-7516/2015/9/023}{{\em JCAP} {\bfseries 1509} no.~09, (2015)
  023},
\href{http://arxiv.org/abs/1504.04276}{{\ttfamily arXiv:1504.04276
  [astro-ph.HE]}}.

\bibitem{Aad:2015baa}
{\bfseries ATLAS} Collaboration, G.~Aad {\em et~al.}, ``{Summary of the ATLAS
  experiments sensitivity to supersymmetry after LHC Run 1 -- interpreted in
  the phenomenological MSSM},''
  \href{http://dx.doi.org/10.1007/JHEP10(2015)134}{{\em JHEP} {\bfseries 10}
  (2015) 134},
\href{http://arxiv.org/abs/1508.06608}{{\ttfamily arXiv:1508.06608 [hep-ex]}}.

\bibitem{Khachatryan:2016nvf}
{\bfseries CMS} Collaboration, V.~Khachatryan {\em et~al.}, ``{Phenomenological
  MSSM interpretation of CMS searches in pp collisions at sqrt(s) = 7 and 8
  TeV},'' \href{http://dx.doi.org/10.1007/JHEP10(2016)129}{{\em JHEP}
  {\bfseries 10} (2016) 129},
\href{http://arxiv.org/abs/1606.03577}{{\ttfamily arXiv:1606.03577 [hep-ex]}}.

\bibitem{Kim:2017mtc}
J.~Kim and J.~McDonald, ``{A Clockwork Higgs Portal Model for Freeze-In Dark
  Matter},''
\href{http://arxiv.org/abs/1709.04105}{{\ttfamily arXiv:1709.04105 [hep-ph]}}.

\bibitem{Kaneta:2016wvf}
K.~Kaneta, H.-S. Lee, and S.~Yun, ``{Portal Connecting Dark Photons and
  Axions},'' \href{http://dx.doi.org/10.1103/PhysRevLett.118.101802}{{\em Phys.
  Rev. Lett.} {\bfseries 118} no.~10, (2017) 101802},
\href{http://arxiv.org/abs/1611.01466}{{\ttfamily arXiv:1611.01466 [hep-ph]}}.

\bibitem{Heikinheimo:2016yds}
M.~Heikinheimo, T.~Tenkanen, K.~Tuominen, and V.~Vaskonen, ``{Observational
  Constraints on Decoupled Hidden Sectors},''
  \href{http://dx.doi.org/10.1103/PhysRevD.96.109902,
  10.1103/PhysRevD.94.063506}{{\em Phys. Rev.} {\bfseries D94} no.~6, (2016)
  063506}, \href{http://arxiv.org/abs/1604.02401}{{\ttfamily arXiv:1604.02401
  [astro-ph.CO]}}.
[Erratum: Phys. Rev.D96,no.10,109902(2017)].

\bibitem{Ayazi:2015jij}
S.~Yaser~Ayazi, S.~M. Firouzabadi, and S.~P. Zakeri, ``{Freeze-in production of
  Fermionic Dark Matter with Pseudo-scalar and Phenomenological Aspects},''
  \href{http://dx.doi.org/10.1088/0954-3899/43/9/095006}{{\em J. Phys.}
  {\bfseries G43} no.~9, (2016) 095006},
\href{http://arxiv.org/abs/1511.07736}{{\ttfamily arXiv:1511.07736 [hep-ph]}}.

\bibitem{Molinaro:2014lfa}
E.~Molinaro, C.~E. Yaguna, and O.~Zapata, ``{FIMP realization of the scotogenic
  model},'' \href{http://dx.doi.org/10.1088/1475-7516/2014/07/015}{{\em JCAP}
  {\bfseries 1407} (2014) 015},
\href{http://arxiv.org/abs/1405.1259}{{\ttfamily arXiv:1405.1259 [hep-ph]}}.

\bibitem{Blennow:2013jba}
M.~Blennow, E.~Fernandez-Martinez, and B.~Zaldivar, ``{Freeze-in through
  portals},'' \href{http://dx.doi.org/10.1088/1475-7516/2014/01/003}{{\em JCAP}
  {\bfseries 1401} (2014) 003},
\href{http://arxiv.org/abs/1309.7348}{{\ttfamily arXiv:1309.7348 [hep-ph]}}.

\bibitem{Dev:2013yza}
P.~S. Bhupal~Dev, A.~Mazumdar, and S.~Qutub, ``{Constraining Non-thermal and
  Thermal properties of Dark Matter},''
  \href{http://dx.doi.org/10.3389/fphy.2014.00026}{{\em Front.in Phys.}
  {\bfseries 2} (2014) 26},
\href{http://arxiv.org/abs/1311.5297}{{\ttfamily arXiv:1311.5297 [hep-ph]}}.

\bibitem{Chu:2013jja}
X.~Chu, Y.~Mambrini, J.~Quevillon, and B.~Zaldivar, ``{Thermal and non-thermal
  production of dark matter via Z'-portal(s)},''
  \href{http://dx.doi.org/10.1088/1475-7516/2014/01/034}{{\em JCAP} {\bfseries
  1401} (2014) 034},
\href{http://arxiv.org/abs/1306.4677}{{\ttfamily arXiv:1306.4677 [hep-ph]}}.

\bibitem{Klasen:2013ypa}
M.~Klasen and C.~E. Yaguna, ``{Warm and cold fermionic dark matter via
  freeze-in},'' \href{http://dx.doi.org/10.1088/1475-7516/2013/11/039}{{\em
  JCAP} {\bfseries 1311} (2013) 039},
\href{http://arxiv.org/abs/1309.2777}{{\ttfamily arXiv:1309.2777 [hep-ph]}}.

\bibitem{Mambrini:2013iaa}
Y.~Mambrini, K.~A. Olive, J.~Quevillon, and B.~Zaldivar, ``{Gauge Coupling
  Unification and Nonequilibrium Thermal Dark Matter},''
  \href{http://dx.doi.org/10.1103/PhysRevLett.110.241306}{{\em Phys. Rev.
  Lett.} {\bfseries 110} no.~24, (2013) 241306},
\href{http://arxiv.org/abs/1302.4438}{{\ttfamily arXiv:1302.4438 [hep-ph]}}.

\bibitem{Yaguna:2011qn}
C.~E. Yaguna, ``{The Singlet Scalar as FIMP Dark Matter},''
  \href{http://dx.doi.org/10.1007/JHEP08(2011)060}{{\em JHEP} {\bfseries 08}
  (2011) 060},
\href{http://arxiv.org/abs/1105.1654}{{\ttfamily arXiv:1105.1654 [hep-ph]}}.

\bibitem{Chu:2011be}
X.~Chu, T.~Hambye, and M.~H.~G. Tytgat, ``{The Four Basic Ways of Creating Dark
  Matter Through a Portal},''
  \href{http://dx.doi.org/10.1088/1475-7516/2012/05/034}{{\em JCAP} {\bfseries
  1205} (2012) 034},
\href{http://arxiv.org/abs/1112.0493}{{\ttfamily arXiv:1112.0493 [hep-ph]}}.

\bibitem{Shakya:2015xnx}
B.~Shakya, ``{Sterile Neutrino Dark Matter from Freeze-In},''
  \href{http://dx.doi.org/10.1142/S0217732316300056}{{\em Mod. Phys. Lett.}
  {\bfseries A31} no.~06, (2016) 1630005},
\href{http://arxiv.org/abs/1512.02751}{{\ttfamily arXiv:1512.02751 [hep-ph]}}.

\bibitem{Merle:2015oja}
A.~Merle and M.~Totzauer, ``{keV Sterile Neutrino Dark Matter from Singlet
  Scalar Decays: Basic Concepts and Subtle Features},''
  \href{http://dx.doi.org/10.1088/1475-7516/2015/06/011}{{\em JCAP} {\bfseries
  1506} (2015) 011},
\href{http://arxiv.org/abs/1502.01011}{{\ttfamily arXiv:1502.01011 [hep-ph]}}.

\bibitem{Nurmi:2015ema}
S.~Nurmi, T.~Tenkanen, and K.~Tuominen, ``{Inflationary Imprints on Dark
  Matter},'' \href{http://dx.doi.org/10.1088/1475-7516/2015/11/001}{{\em JCAP}
  {\bfseries 1511} no.~11, (2015) 001},
\href{http://arxiv.org/abs/1506.04048}{{\ttfamily arXiv:1506.04048
  [astro-ph.CO]}}.

\bibitem{Pandey:2017quk}
M.~Pandey, D.~Majumdar, and K.~P. Modak, ``{Two Component Feebly Interacting
  Massive Particle (FIMP) Dark Matter},''
\href{http://arxiv.org/abs/1709.05955}{{\ttfamily arXiv:1709.05955 [hep-ph]}}.

\bibitem{Arcadi:2016dbl}
G.~Arcadi, P.~Ghosh, Y.~Mambrini, and M.~Pierre, ``{Re-opening dark matter
  windows compatible with a diphoton excess},''
  \href{http://dx.doi.org/10.1088/1475-7516/2016/07/005}{{\em JCAP} {\bfseries
  1607} no.~07, (2016) 005},
\href{http://arxiv.org/abs/1603.05601}{{\ttfamily arXiv:1603.05601 [hep-ph]}}.

\bibitem{Benakli:2017whb}
K.~Benakli, Y.~Chen, E.~Dudas, and Y.~Mambrini, ``{Minimal model of gravitino
  dark matter},'' \href{http://dx.doi.org/10.1103/PhysRevD.95.095002}{{\em
  Phys. Rev.} {\bfseries D95} no.~9, (2017) 095002},
\href{http://arxiv.org/abs/1701.06574}{{\ttfamily arXiv:1701.06574 [hep-ph]}}.

\bibitem{McDonald:2001vt}
J.~McDonald, ``{Thermally generated gauge singlet scalars as selfinteracting
  dark matter},'' \href{http://dx.doi.org/10.1103/PhysRevLett.88.091304}{{\em
  Phys. Rev. Lett.} {\bfseries 88} (2002) 091304},
\href{http://arxiv.org/abs/hep-ph/0106249}{{\ttfamily arXiv:hep-ph/0106249
  [hep-ph]}}.

\bibitem{Hall:2009bx}
L.~J. Hall, K.~Jedamzik, J.~March-Russell, and S.~M. West, ``{Freeze-In
  Production of FIMP Dark Matter},''
  \href{http://dx.doi.org/10.1007/JHEP03(2010)080}{{\em JHEP} {\bfseries 03}
  (2010) 080},
\href{http://arxiv.org/abs/0911.1120}{{\ttfamily arXiv:0911.1120 [hep-ph]}}.

\bibitem{Elahi:2014fsa}
F.~Elahi, C.~Kolda, and J.~Unwin, ``{UltraViolet Freeze-in},''
  \href{http://dx.doi.org/10.1007/JHEP03(2015)048}{{\em JHEP} {\bfseries 03}
  (2015) 048},
\href{http://arxiv.org/abs/1410.6157}{{\ttfamily arXiv:1410.6157 [hep-ph]}}.

\bibitem{McDonald:2015ljz}
J.~McDonald, ``{Warm Dark Matter via Ultra-Violet Freeze-In: Reheating
  Temperature and Non-Thermal Distribution for Fermionic Higgs Portal Dark
  Matter},'' \href{http://dx.doi.org/10.1088/1475-7516/2016/08/035}{{\em JCAP}
  {\bfseries 1608} no.~08, (2016) 035},
\href{http://arxiv.org/abs/1512.06422}{{\ttfamily arXiv:1512.06422 [hep-ph]}}.

\bibitem{Mambrini:2015vna}
Y.~Mambrini, N.~Nagata, K.~A. Olive, J.~Quevillon, and J.~Zheng, ``{Dark matter
  and gauge coupling unification in nonsupersymmetric SO(10) grand unified
  models},'' \href{http://dx.doi.org/10.1103/PhysRevD.91.095010}{{\em Phys.
  Rev.} {\bfseries D91} no.~9, (2015) 095010},
\href{http://arxiv.org/abs/1502.06929}{{\ttfamily arXiv:1502.06929 [hep-ph]}}.

\bibitem{vll}
G.~B\'elanger, A.~Goudelis, A.~Pukhov, and B.~Zald\'ivar.
\newblock {in progress}.

\bibitem{Barducci:2016pcb}
D.~Barducci, G.~Belanger, J.~Bernon, F.~Boudjema, J.~Da~Silva, S.~Kraml,
  U.~Laa, and A.~Pukhov, ``{Collider limits on new physics within
  micrOMEGAs4.3},''
\href{http://arxiv.org/abs/1606.03834}{{\ttfamily arXiv:1606.03834 [hep-ph]}}.

\bibitem{Belanger:2006is}
G.~Belanger, F.~Boudjema, A.~Pukhov, and A.~Semenov, ``{MicrOMEGAs 2.0: A
  Program to calculate the relic density of dark matter in a generic model},''
  \href{http://dx.doi.org/10.1016/j.cpc.2006.11.008}{{\em Comput. Phys.
  Commun.} {\bfseries 176} (2007) 367--382},
\href{http://arxiv.org/abs/hep-ph/0607059}{{\ttfamily arXiv:hep-ph/0607059
  [hep-ph]}}.

\bibitem{Belanger:2001fz}
G.~Belanger, F.~Boudjema, A.~Pukhov, and A.~Semenov, ``{MicrOMEGAs: A Program
  for calculating the relic density in the MSSM},''
  \href{http://dx.doi.org/10.1016/S0010-4655(02)00596-9}{{\em Comput. Phys.
  Commun.} {\bfseries 149} (2002) 103--120},
\href{http://arxiv.org/abs/hep-ph/0112278}{{\ttfamily arXiv:hep-ph/0112278
  [hep-ph]}}.

\bibitem{Gondolo:2004sc}
P.~Gondolo, J.~Edsjo, P.~Ullio, L.~Bergstrom, M.~Schelke, and E.~A. Baltz,
  ``{DarkSUSY: Computing supersymmetric dark matter properties numerically},''
  \href{http://dx.doi.org/10.1088/1475-7516/2004/07/008}{{\em JCAP} {\bfseries
  0407} (2004) 008},
\href{http://arxiv.org/abs/astro-ph/0406204}{{\ttfamily arXiv:astro-ph/0406204
  [astro-ph]}}.

\bibitem{Arbey:2011zz}
A.~Arbey and F.~Mahmoudi, ``{SuperIso Relic v3.0: A program for calculating
  relic density and flavour physics observables: Extension to NMSSM},''
\href{http://dx.doi.org/10.1016/j.cpc.2011.03.019}{{\em Comput. Phys. Commun.}
  {\bfseries 182} (2011) 1582--1583}.

\bibitem{Backovic:2013dpa}
M.~Backovic, K.~Kong, and M.~McCaskey, ``{MadDM v.1.0: Computation of Dark
  Matter Relic Abundance Using MadGraph5},''
  \href{http://dx.doi.org/10.1016/j.dark.2014.04.001}{{\em Physics of the Dark
  Universe} {\bfseries 5-6} (2014) 18--28},
\href{http://arxiv.org/abs/1308.4955}{{\ttfamily arXiv:1308.4955 [hep-ph]}}.

\bibitem{Bernal:2017kxu}
N.~Bernal, M.~Heikinheimo, T.~Tenkanen, K.~Tuominen, and V.~Vaskonen, ``{The
  Dawn of FIMP Dark Matter: A Review of Models and Constraints},''
\href{http://arxiv.org/abs/1706.07442}{{\ttfamily arXiv:1706.07442 [hep-ph]}}.

\bibitem{Merle:2013wta}
A.~Merle, V.~Niro, and D.~Schmidt, ``{New Production Mechanism for keV Sterile
  Neutrino Dark Matter by Decays of Frozen-In Scalars},''
  \href{http://dx.doi.org/10.1088/1475-7516/2014/03/028}{{\em JCAP} {\bfseries
  1403} (2014) 028},
\href{http://arxiv.org/abs/1306.3996}{{\ttfamily arXiv:1306.3996 [hep-ph]}}.

\bibitem{An:2014twa}
H.~An, M.~Pospelov, J.~Pradler, and A.~Ritz, ``{Direct Detection Constraints on
  Dark Photon Dark Matter},''
  \href{http://dx.doi.org/10.1016/j.physletb.2015.06.018}{{\em Phys. Lett.}
  {\bfseries B747} (2015) 331--338},
\href{http://arxiv.org/abs/1412.8378}{{\ttfamily arXiv:1412.8378 [hep-ph]}}.

\bibitem{Essig:2013goa}
R.~Essig, E.~Kuflik, S.~D. McDermott, T.~Volansky, and K.~M. Zurek,
  ``{Constraining Light Dark Matter with Diffuse X-Ray and Gamma-Ray
  Observations},'' \href{http://dx.doi.org/10.1007/JHEP11(2013)193}{{\em JHEP}
  {\bfseries 11} (2013) 193},
\href{http://arxiv.org/abs/1309.4091}{{\ttfamily arXiv:1309.4091 [hep-ph]}}.

\bibitem{Heikinheimo:2018duk}
M.~Heikinheimo, T.~Tenkanen, and K.~Tuominen, ``{Prospects for indirect
  detection of frozen-in dark matter},''
\href{http://arxiv.org/abs/1801.03089}{{\ttfamily arXiv:1801.03089 [hep-ph]}}.

\bibitem{Arcadi:2014tsa}
G.~Arcadi, L.~Covi, and F.~Dradi, ``{LHC prospects for minimal decaying Dark
  Matter},'' \href{http://dx.doi.org/10.1088/1475-7516/2014/10/063}{{\em JCAP}
  {\bfseries 1410} no.~10, (2014) 063},
\href{http://arxiv.org/abs/1408.1005}{{\ttfamily arXiv:1408.1005 [hep-ph]}}.

\bibitem{Aaboud:2016uth}
{\bfseries ATLAS} Collaboration, M.~Aaboud {\em et~al.}, ``{Search for heavy
  long-lived charged $R$-hadrons with the ATLAS detector in 3.2 fb$^{-1}$ of
  proton--proton collision data at $\sqrt{s} = 13$ TeV},''
  \href{http://dx.doi.org/10.1016/j.physletb.2016.07.042}{{\em Phys. Lett.}
  {\bfseries B760} (2016) 647--665},
\href{http://arxiv.org/abs/1606.05129}{{\ttfamily arXiv:1606.05129 [hep-ex]}}.

\bibitem{Khachatryan:2016sfv}
{\bfseries CMS} Collaboration, V.~Khachatryan {\em et~al.}, ``{Search for
  long-lived charged particles in proton-proton collisions at $\sqrt s=$ 13
  TeV},'' \href{http://dx.doi.org/10.1103/PhysRevD.94.112004}{{\em Phys. Rev.}
  {\bfseries D94} no.~11, (2016) 112004},
\href{http://arxiv.org/abs/1609.08382}{{\ttfamily arXiv:1609.08382 [hep-ex]}}.

\bibitem{Ghosh:2017vhe}
A.~Ghosh, T.~Mondal, and B.~Mukhopadhyaya, ``{Heavy stable charged tracks as
  signatures of non-thermal dark matter at the LHC : a study in some
  non-supersymmetric scenarios},''
\href{http://arxiv.org/abs/1706.06815}{{\ttfamily arXiv:1706.06815 [hep-ph]}}.

\bibitem{Kolb:1990vq}
E.~W. Kolb and M.~S. Turner, ``{The Early Universe},''
{\em Front. Phys.} {\bfseries 69} (1990) 1--547.

\bibitem{Gondolo:1990dk}
P.~Gondolo and G.~Gelmini, ``{Cosmic abundances of stable particles: Improved
  analysis},''
\href{http://dx.doi.org/10.1016/0550-3213(91)90438-4}{{\em Nucl. Phys.}
  {\bfseries B360} (1991) 145--179}.

\bibitem{Belanger:2014vza}
G.~Belanger, F.~Boudjema, A.~Pukhov, and A.~Semenov, ``{micrOMEGAs4.1: two dark
  matter candidates},'' \href{http://dx.doi.org/10.1016/j.cpc.2015.03.003}{{\em
  Comput. Phys. Commun.} {\bfseries 192} (2015) 322--329},
\href{http://arxiv.org/abs/1407.6129}{{\ttfamily arXiv:1407.6129 [hep-ph]}}.

\bibitem{Abercrombie:2015wmb}
D.~Abercrombie {\em et~al.}, ``{Dark Matter Benchmark Models for Early LHC
  Run-2 Searches: Report of the ATLAS/CMS Dark Matter Forum},''
\href{http://arxiv.org/abs/1507.00966}{{\ttfamily arXiv:1507.00966 [hep-ex]}}.

\bibitem{Chung:1998zb}
D.~J.~H. Chung, E.~W. Kolb, and A.~Riotto, ``{Superheavy dark matter},''
  \href{http://dx.doi.org/10.1103/PhysRevD.59.023501}{{\em Phys. Rev.}
  {\bfseries D59} (1999) 023501},
\href{http://arxiv.org/abs/hep-ph/9802238}{{\ttfamily arXiv:hep-ph/9802238
  [hep-ph]}}.

\bibitem{Garcia:2017tuj}
M.~A.~G. Garcia, Y.~Mambrini, K.~A. Olive, and M.~Peloso, ``{Enhancement of the
  Dark Matter Abundance Before Reheating: Applications to Gravitino Dark
  Matter},'' \href{http://dx.doi.org/10.1103/PhysRevD.96.103510}{{\em Phys.
  Rev.} {\bfseries D96} no.~10, (2017) 103510},
\href{http://arxiv.org/abs/1709.01549}{{\ttfamily arXiv:1709.01549 [hep-ph]}}.

\bibitem{Chen:2017kvz}
S.-L. Chen and Z.~Kang, ``{On UltraViolet Freeze-in Dark Matter during
  Reheating},''
\href{http://arxiv.org/abs/1711.02556}{{\ttfamily arXiv:1711.02556 [hep-ph]}}.

\bibitem{Bernal:2015ova}
N.~Bernal, X.~Chu, C.~Garcia-Cely, T.~Hambye, and B.~Zaldivar, ``{Production
  Regimes for Self-Interacting Dark Matter},''
  \href{http://dx.doi.org/10.1088/1475-7516/2016/03/018}{{\em JCAP} {\bfseries
  1603} no.~03, (2016) 018},
\href{http://arxiv.org/abs/1510.08063}{{\ttfamily arXiv:1510.08063 [hep-ph]}}.

\bibitem{Bernal:2015xba}
N.~Bernal and X.~Chu, ``{$\mathbb {Z}_2$ SIMP Dark Matter},''
  \href{http://dx.doi.org/10.1088/1475-7516/2016/01/006}{{\em JCAP} {\bfseries
  1601} (2016) 006},
\href{http://arxiv.org/abs/1510.08527}{{\ttfamily arXiv:1510.08527 [hep-ph]}}.

\bibitem{Alloul:2013bka}
A.~Alloul, N.~D. Christensen, C.~Degrande, C.~Duhr, and B.~Fuks, ``{FeynRules
  2.0 - A complete toolbox for tree-level phenomenology},''
  \href{http://dx.doi.org/10.1016/j.cpc.2014.04.012}{{\em Comput. Phys.
  Commun.} {\bfseries 185} (2014) 2250--2300},
\href{http://arxiv.org/abs/1310.1921}{{\ttfamily arXiv:1310.1921 [hep-ph]}}.

\bibitem{Belyaev:2012qa}
A.~Belyaev, N.~D. Christensen, and A.~Pukhov, ``{CalcHEP 3.4 for collider
  physics within and beyond the Standard Model},''
  \href{http://dx.doi.org/10.1016/j.cpc.2013.01.014}{{\em Comput. Phys.
  Commun.} {\bfseries 184} (2013) 1729--1769},
\href{http://arxiv.org/abs/1207.6082}{{\ttfamily arXiv:1207.6082 [hep-ph]}}.

\bibitem{micro-hp}
\url{https://lapth.cnrs.fr/micromegas}.

\bibitem{Bae:2017dpt}
K.~J. Bae, A.~Kamada, S.~P. Liew, and K.~Yanagi, ``{Light Axinos from
  Freeze-in: production processes, phase space distributions, and Ly-$\alpha$
  constraints},''
\href{http://arxiv.org/abs/1707.06418}{{\ttfamily arXiv:1707.06418 [hep-ph]}}.

\end{thebibliography}\endgroup

\end{document}